\documentclass[10pt,conference]{IEEEtran}
\IEEEoverridecommandlockouts
\usepackage{cite}
\usepackage{amsmath,amssymb,amsfonts}
\usepackage{algorithmic}
\usepackage{graphicx}
\usepackage{textcomp}
\usepackage{xcolor}
\usepackage[ruled,linesnumbered]{algorithm2e}
\usepackage{subfigure}
\usepackage{multirow} 
\usepackage{booktabs}
\usepackage{url}

\usepackage{bbding}
\usepackage[final]{changes}

\usepackage[table]{xcolor} 
\definecolor{pos}{RGB}{210,245,239}
\definecolor{neg}{RGB}{253,213,210}
\usepackage{tcolorbox}
\definechangesauthor[name={Dcm}, color=red]{dcm.}
\def\BibTeX{{\rm B\kern-.05em{\sc i\kern-.025em b}\kern-.08em
    T\kern-.1667em\lower.7ex\hbox{E}\kern-.125emX}}
\begin{document}

\title{United We Stand: Towards End-to-End Log-based Fault Diagnosis via Interactive Multi-Task Learning\\
\thanks{\IEEEauthorrefmark{1} Corresponding Authors.}
\thanks{\IEEEauthorrefmark{5} These authors contributed equally to this work.}
}

% \IEEEauthorblockA{\textit{dept. name of organization (of Aff.)} \\X
% \and
% \and
% \and
% \and
% \and

\author{
    \IEEEauthorblockN{
        Minghua He\IEEEauthorrefmark{2}\IEEEauthorrefmark{5},
        Chiming Duan\IEEEauthorrefmark{2}\IEEEauthorrefmark{5},
        Pei Xiao\IEEEauthorrefmark{2}\IEEEauthorrefmark{5},
        Tong Jia\IEEEauthorrefmark{2}\IEEEauthorrefmark{3}\IEEEauthorrefmark{1},
        Siyu Yu\IEEEauthorrefmark{2}, \\
        Lingzhe Zhang\IEEEauthorrefmark{2},
        Weijie Hong\IEEEauthorrefmark{2},
        Jin Han\IEEEauthorrefmark{4},
        Yifan Wu\IEEEauthorrefmark{2},
        Ying Li\IEEEauthorrefmark{2},
        and Gang Huang\IEEEauthorrefmark{2}\IEEEauthorrefmark{3}
    }
    \IEEEauthorblockA
    {
    \IEEEauthorrefmark{2}Peking University, Beijing, China\\ 
    \IEEEauthorrefmark{3}National Key Laboratory of Data Space Technology and System, Beijing, China\\ 
    \IEEEauthorrefmark{4}ZTE Corporation, Beijing, China \\
    \{hemh2120, duanchiming, xiaopei, yusiyu, zhang.lingzhe\}@stu.pku.edu.cn \\ 
    \{jia.tong, yifanwu, li.ying, hg\}@pku.edu.cn, han.jing28@zte.com.cn
    }
}

\maketitle

\begin{abstract}
Log-based fault diagnosis is essential for maintaining software system availability. However, existing fault diagnosis methods are built using a task-independent manner, which fails to bridge the gap between anomaly detection and root cause localization in terms of data form and diagnostic objectives, resulting in three major issues: 1) Diagnostic bias accumulates in the system; 2) System deployment relies on expensive monitoring data; 3) The collaborative relationship between diagnostic tasks is overlooked. Facing this problems, we propose a novel end-to-end log-based fault diagnosis method, Chimera, whose key idea is to achieve end-to-end fault diagnosis through bidirectional interaction and knowledge transfer between anomaly detection and root cause localization. Chimera is based on interactive multi-task learning, carefully designing interaction strategies between anomaly detection and root cause localization at the data, feature, and diagnostic result levels, thereby achieving both sub-tasks interactively within a unified end-to-end framework. Evaluation on two public datasets and one industrial dataset shows that Chimera outperforms existing methods in both anomaly detection and root cause localization, achieving improvements of over 2.92\%$\sim$5.00\% and 19.01\%$\sim$37.09\%, respectively. It has been successfully deployed in production, serving an industrial cloud platform.
\end{abstract}


\section{Introduction}
Software systems (e.g., web servers) are becoming increasingly large and complex and are subject to more failures \cite{niu2023locating, he2025execoder, lin2023edits, chen2025warriormath, li2023elastic, luo2021ntam, llmelog, huang2024demystifying}. 
Software providers must find rapid and effective methods to manage exceptions in order to enhance system reliability. \replaced[id=dcm.]{Faults can propagate across servers, causing anomalies within the system and ultimately leading to failures of the entire system \cite{chen2024lara, tan2024air, yu2024supervised, dai2021sdfvae, huang2022semi}.}{In a real-world context where availability is crucial for customer experience, service providers must find methods to quickly and effectively manage anomalous events to enhance system reliability.} Fault diagnosis requires the timely \replaced[id=dcm.]{detect the anomalies} { identification of abnormal states} within the system and the accurate localization of \replaced[id=dcm.]{root causes}{root cause events} that trigger failures \cite{wong2016survey, zhang2024failure, jiang2011efficient}, which is crucial for Site Reliability Engineers (SREs) to establish \deleted[id=dcm.]{failure} recovery strategies and maintain the reliability of software systems.

\begin{figure}[htbp]
	\centering  
	\subfigbottomskip=5pt 
	\subfigcapskip=5pt 
	\subfigure[\textbf{The Existing Task-Independent Paradigm}. Initially, the labeled log sequences and log entries are inputted, and the detector and localizer are trained independently. The two are integrated into a fault diagnosis network, leading to issues such as diagnostic bias accumulates.]{
		\includegraphics[width=0.9\linewidth]{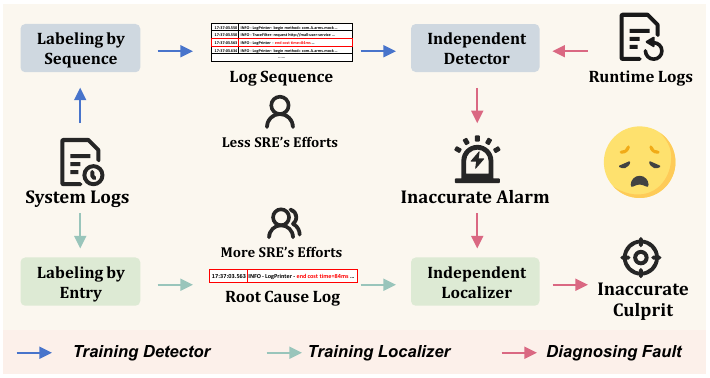}\label{teaser1}}
	\subfigure[\textbf{The Proposed Task-Interactive Paradigm}. The labeled log sequence is used to train the detector and localizer interactively in an end-to-end manner, resulting in excellent diagnostic performance.]{
		\includegraphics[width=0.9\linewidth]{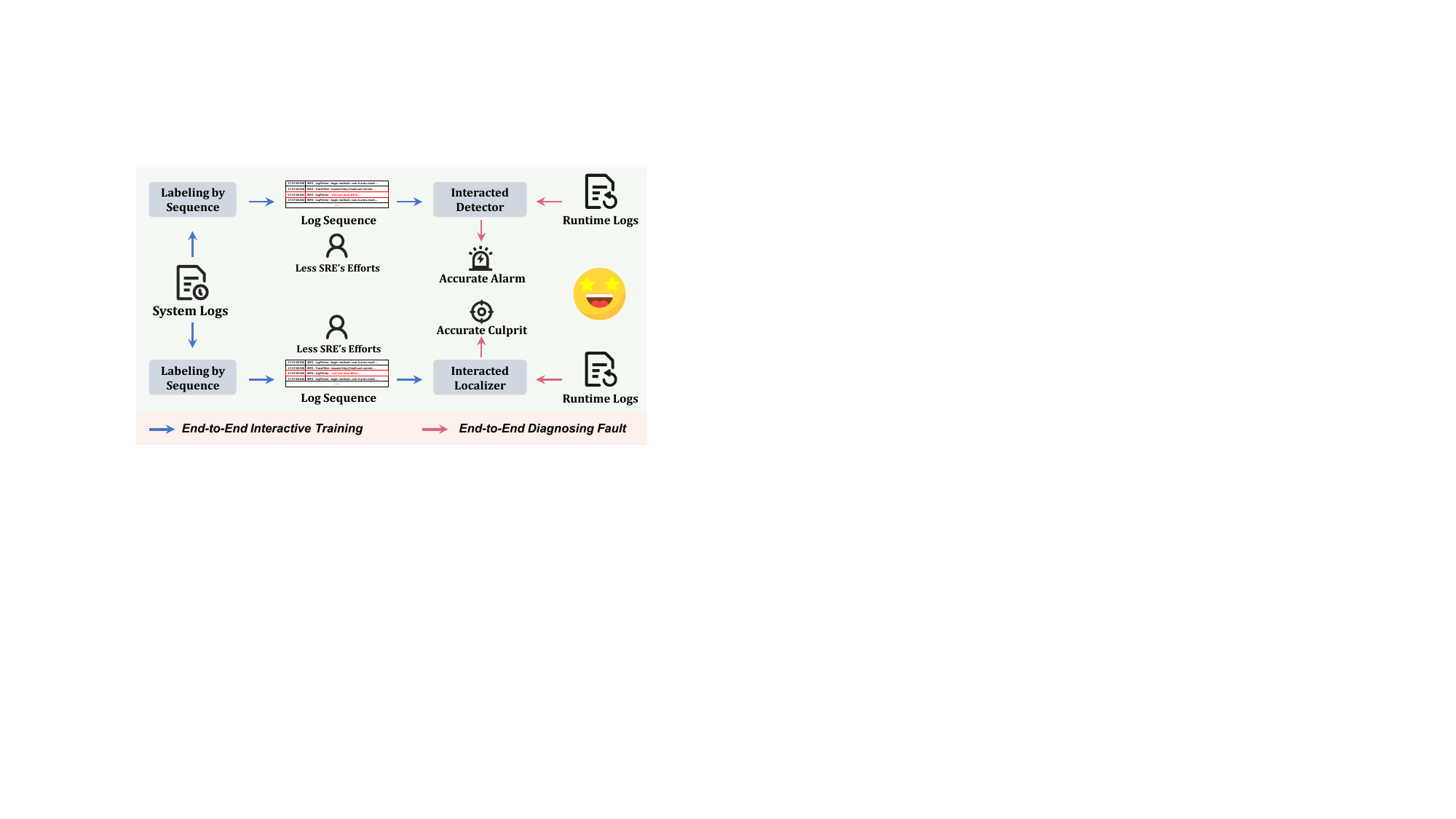}\label{teaser2}}
  % \captionsetup{skip=2pt}
  \caption{Two different fault diagnosis deployment paradigms. The proposed method achieves end-to-end fault diagnosis through bidirectional interaction between anomaly detection and root cause localization.} 
  \label{teaser}
  \vspace{-0.4cm}
\end{figure}

Logs record the system's operational state and are an essential source of information for \replaced[id=dcm.]{fault}{failure} diagnosis. In recent years, there has been significant research on log-based fault diagnosis \cite{zhang2019robust,xia2019bugidentifier, sun2025exploring, zhao2025few, zhang2025survey, reidemeister2010identifying,tong2016approach,yang2021semi,du2017deeplog,meng2019loganomaly,jiang2025l4,huang2025no,zhang2024metalog,jia2024hilogx, zhang2021cloudrca}. Current log-based fault diagnosis techniques \cite{wittkopp2024logrca, hou2021diagnosing, huo2023evlog, zhang2021onion, he2018identifying, amar2019mining, lee2023eadro} generally consist of two separate stages: anomaly detection and root cause localization. An \replaced[id=dcm.]{log entry}{individual log} captures specific operational events of the system and can uncover the root cause at the time of a failure \cite{notaro2023logrule, wang2020root}. Additionally,\replaced[id=dcm.]{ log sequences, with multiple log entries organized sequentially, }{multiple logs organized sequentially} provide a complete context of operations, reflecting the system's operational state more thoroughly \cite{yang2018nanolog, luo2018troubleshooting}. Thus, in the anomaly detection phase, current methods \cite{zhang2019robust,xia2019bugidentifier, duan2023aclog,tong2016approach} employ sequential neural networks to extract the system's operational state from log sequences for anomaly detection. After an anomaly is detected, the subsequent process involves localizing the root cause. The purpose of root cause localization is to identify the \replaced[id=dcm.]{fault}{culprit} that caused the failure and link it to specific log entries \cite{yu2023nezha, rosenberg2020spectrum}. By examining these root cause logs, SREs can comprehend the specific operational events that led to the failure, enabling them to quickly formulate relevant recovery strategies \cite{zhang2021onion, jia2021logflash, shi2023serverrca}.

Despite significant efforts, achieving accurate fault diagnosis still faces many challenges, as shown in figure \ref{teaser}. 
First, diagnostic bias accumulates within the system. Log-based fault diagnosis is divided into two stages: first, anomaly detection, followed by root cause localization. Existing methods execute these two stages independently, leading to inaccurate anomaly detection that severely affects the effectiveness of root cause localization.
Secondly, system deployment relies on costly monitoring data. Existing methods require \added[id=dcm.]{not noly} the provision of anomalous log sequences \added[id=dcm.]{but also} root cause logs to separately train the detector and the localizer. However, collecting substantial amounts of these two different types of log data requires significant human \replaced[id=dcm.]{efforts}{effort}, making it nearly impractical. 
Finally, the collaborative relationship between diagnostic tasks is overlooked. Anomaly detection and root cause localization are interrelated tasks, and executing these two tasks independently neglects the rich transfer of information between them, which results in suboptimal diagnosis.

In our view, the root cause of these issues is that anomaly detection and root cause localization rely on different \replaced[id=dcm.]{forms}{types} of data for training and have distinct diagnostic objectives for faults. The anomaly detection uses anomalous log sequences as training data, with the diagnostic objective of determining whether a fault has occurred in the system. The root cause localization utilizes root cause logs as training data, with the objective of localizing the \replaced[id=dcm.]{fault}{culprit} that caused the \replaced[id=dcm.]{failure}{fault}. Simply piecing together expert models to construct a fault diagnosis system cannot effectively bridge the gap between the two in terms of data forms and diagnostic objectives, nor can it handle fault diagnosis within a unified framework. 

Addressing these issues requires bridging the gap between the two sub-tasks in terms of data forms and diagnostic objectives, and constructing an end-to-end fault diagnosis system. Our approach is to implement interactive multi-task learning between the two sub-tasks. Our key insight is that there is a strong mutual implication between the anomaly detection task and the root cause localization task, and their bidirectional interaction and knowledge transfer can bridge the gaps in data forms and diagnostic objectives.

In this paper, we introduce \textbf{Chimera}, an innovative end-to-end log-based fault diagnosis method designed to achieve anomaly detection and root cause localization through interaction at three levels: data, features, and diagnostic results, within a unified framework. Chimera first interacts at the data level to bridge the gap in data forms. It leverages the principle of multiple-instance learning \cite{sultani2018real}\cite{midlog} to develop a localizer trained only on anomalous log sequences. The key idea is to continuously compare normal log sequences with anomalous log sequences to infer the root cause from the anomalous log sequences. Subsequently, Chimera interacts at the feature and diagnostic result levels to bridge the gap in diagnostic objectives. At the feature level, Chimera employs disentanglement learning to learn interactive log representations for the two tasks. At the diagnostic result level, Chimera aligns the diagnostic results of the two tasks based on mutual information theory. With a meticulously designed interactive multi-task learning strategy, Chimera can perform fault diagnosis in an end-to-end manner, avoiding potential error propagation. Moreover, the bidirectional interaction between the sub-tasks fully leverages the diagnostic information, resulting in more effective fault diagnosis.

To assess the effectiveness of Chimera, we conducted extensive experiments on two widely used public datasets, BGL and Thunderbird \cite{oliner2007supercomputers}, as well as an industrial dataset, System A. The experimental results indicate that, in the anomaly detection task, Chimera achieved an average F1 score exceeding 90\%, outperforming existing task-independent methods by more than 2.92\% to 5.00\%. In the root cause localization task, compared to existing task-independent methods, Chimera achieved an advantage of over 19.01\% to 37.09\% across the four evaluated metrics. The experiments demonstrated that the bidirectional interaction and knowledge transfer between anomaly detection and root cause localization contribute to more effective fault diagnosis. In summary, the contributions of this paper are as follows:

\begin{itemize}
    \item We proposed an end-to-end log-based fault diagnosis system, Chimera, which achieves anomaly detection and root cause localization interactively within a unified framework through carefully designed interaction strategies.
    \item We designed a sequence-driven localizer based on the principle of multiple-instance learning, which utilizes only anomalous log sequences for training without the need for root cause logs.
    \item We proposed a multi-task interaction strategy based on disentanglement learning and mutual information theory, which facilitates interaction at the feature and diagnostic result levels, effectively promoting knowledge transfer between tasks.
    \item Evaluations on two public datasets and one industrial dataset demonstrate the effectiveness of our approach.
\end{itemize}

\section{Background}
\subsection{Log Terminology}
\begin{figure}[htbp]
\centerline{
\includegraphics[width=0.9\linewidth]{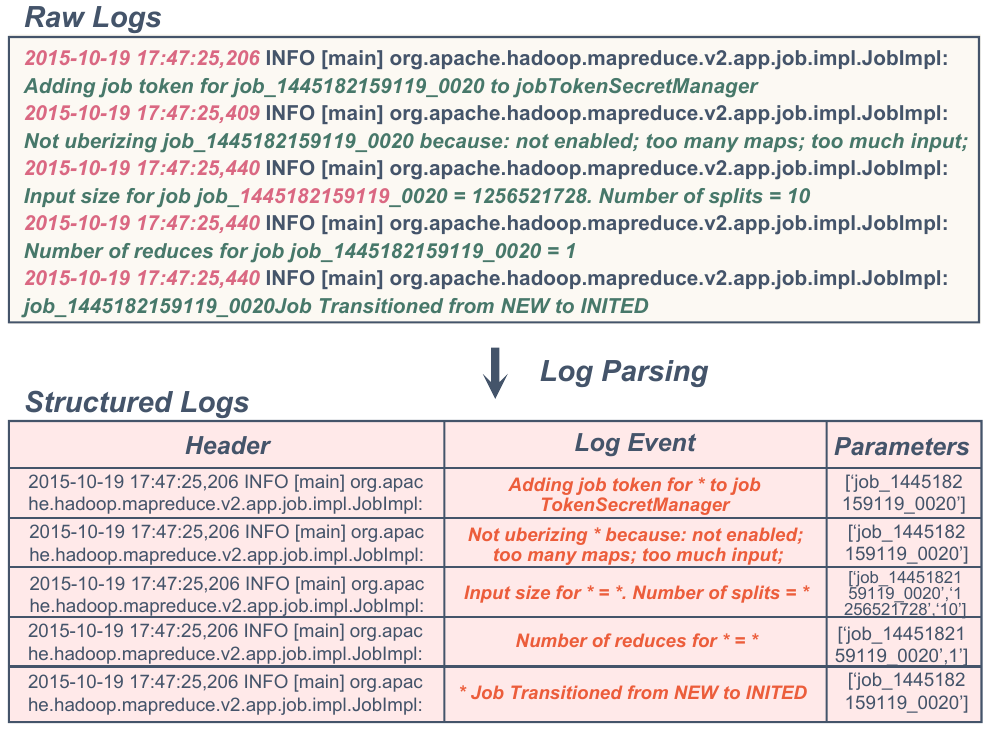}
}
\caption{Examples of key terminologies in log analysis. The red section in raw logs represents timestamps, while the green section represents log contents.}
\label{log_example}
\vspace{-0.4cm}
\end{figure}

In figure \ref{log_example}, we present an example of log data extracted from the publicly available Hadoop dataset, collected from a distributed system. Log messages represent unstructured text generated by logging statements within the system (e.g., logging.info()), containing both log events and parameters. Log events (e.g., "Number of reduces for") form the constant part of log messages, consisting of text strings, and serve as abstractions in log analysis. Log parameters make up the variable part of log messages, recording specific system attributes during runtime (e.g., job ID). Log parsing, typically the initial step in log analysis, separates log events from log parameters in unstructured log messages. An event sequence consists of a sequence of log events that document the execution flow of a particular task, commonly organized by job ID or timestamp.

\subsection{Log-based Fault Diagnosis}
In this paper, our objective is to perform fault diagnosis of software systems based on system logs. Consider a large software system, where we have a log sequence defined as $\mathbf{x} = [l_1, l_2, \ldots, l_n]$ within an observation window of length $n$, where $l _i$ represents the log observed at each time point. Our work seeks to establish an end-to-end framework that processes the input log sequence through a unified model, outputs the system's operational status, and additionally outputs relevant root cause logs when anomalies are detected. Specifically: 1) Given the observed log sequence $x$, the detector $D$ predicts the presence of anomalies, with the result $y$ represented as $0$ (normal) or $1$ (anomalous), i.e., $y = D(\mathbf{x}) \in \{0, 1 \}$. 2) If $y = 1$, the localizer $L$ identifies the probability of each log in the input sequence being the culprit, represented as $L(x) = [p_1, p_2, ..., p_n],  p_i \in [0, 1]$.

\section{Empirical Study}
This section presents an empirical study of existing log-based fault diagnosis methods based on the system log dataset Thunderbird \cite{oliner2007supercomputers}. We manually examined the unsatisfactory performance of existing methods to provide motivational support for the proposed method. Details regarding baselines, metrics, and the dataset can be found in Section \ref{sec:setup}.

% \subsection{Baselines}
% To comprehensively investigate the unsatisfactory performance of existing log-based fault diagnosis methods, we select the most advanced log-based fault diagnosis methods for study, specifically including: 

% \begin{itemize}
%     \item \textbf{RobustFlagger} \cite{amar2019mining}\cite{zhang2019robust}: A fault diagnosis method composed of the advanced log-based anomaly detection method LogRobust \cite{zhang2019robust} and the log-based root cause localization method LogFaultFlagger \cite{amar2019mining} in a pipeline manner.
%     \item \textbf{LogRCA} \cite{wittkopp2024logrca}: The advanced log-based fault diagnosis method that utilizes recurrent neural networks to mine root causes from historical log sequences.
%     \item \textbf{Eadro} \cite{lee2023eadro}: The most advanced multi-modal fault diagnosis method. To ensure fairness in comparison, this paper removes the contributions of metric and trace, using only logs for fault diagnosis of the software system.
%     \item \textbf{AFAFormer} \cite{duan2023afalog}: A fault diagnosis method based on the advanced log-based anomaly detection method AFALog \cite{duan2023afalog}. It utilizes labeled log sequences to train a transformer network for anomaly detection and employs the multi-head attention scores for root cause localization.
% \end{itemize}

\subsection{How effective are existing methods in addressing diagnostic bias?}\label{sec:bias}

Log-based fault diagnosis are divided into two stages: 1) anomaly detection is performed first, and 2) if a fault is detected, further root cause localization is conducted. If the detector yields inaccurate results, these inaccuracies will also be passed on to the localizer. For instance, if the detector reports a false positive, the result from the localizer will also be a false positive. We define this phenomenon as diagnostic bias, which is an unavoidable noise in fault diagnosis.

To further investigate the impact of diagnostic bias in fault diagnosis, we compared the performance of different methods in 1) theoretical anomaly detection settings (referred to as theoretical scores) and 2) actual anomaly detection settings (referred to as actual scores). In the theoretical anomaly detection setting, all system anomalies are passed to the localizer. In the actual anomaly detection setting, we only pass the system anomalies detected by the detector to the localizer. We consider the difference in localizer performance under the two settings as diagnostic bias, with results shown in table \ref{empirical_bias}. 

% Table generated by Excel2LaTeX from sheet 'Add_res'
\begin{table}[htbp]
  \centering
  \caption{The existing methods are affected by diagnostic bias when performing fault diagnosis on the Thunderbird dataset.}
    \begin{tabular}{rrrrr}
    \hline
    \multicolumn{1}{c}{\textbf{Metric}} & \multicolumn{1}{c}{\textbf{Model}} & \multicolumn{1}{c}{\textbf{Theoretical}} & \multicolumn{1}{c}{\textbf{Actual}} & \multicolumn{1}{c}{\textbf{Bias}} \\
    \hline
    \multicolumn{1}{c}{\multirow{5}[4]{*}{HR@1}} & \multicolumn{1}{c}{RobustFlagger} & \multicolumn{1}{c}{19.83} & \multicolumn{1}{c}{13.78} & \multicolumn{1}{c}{-6.05} \\
          & \multicolumn{1}{c}{LogRCA} & \multicolumn{1}{c}{21.00} & \multicolumn{1}{c}{19.05} & \multicolumn{1}{c}{-1.95} \\
          & \multicolumn{1}{c}{AFAFormer} & \multicolumn{1}{c}{12.55} & \multicolumn{1}{c}{11.68} & \multicolumn{1}{c}{-0.87} \\
          & \multicolumn{1}{c}{Eadro} & \multicolumn{1}{c}{28.41} & \multicolumn{1}{c}{14.59} & \multicolumn{1}{c}{-13.82} \\
\cmidrule{2-5}          & \multicolumn{1}{c}{Average} & \multicolumn{1}{c}{20.45} & \multicolumn{1}{c}{14.78} & \multicolumn{1}{c}{\cellcolor{pos}\textbf{-5.67}} \\
    \hline
    \multicolumn{1}{c}{\multirow{5}[4]{*}{HR@3}} & \multicolumn{1}{c}{RobustFlagger} & \multicolumn{1}{c}{39.21} & \multicolumn{1}{c}{22.81} & \multicolumn{1}{c}{-16.40} \\
          & \multicolumn{1}{c}{LogRCA} & \multicolumn{1}{c}{45.21} & \multicolumn{1}{c}{38.50} & \multicolumn{1}{c}{-6.71} \\
          & \multicolumn{1}{c}{AFAFormer} & \multicolumn{1}{c}{47.27} & \multicolumn{1}{c}{39.68} & \multicolumn{1}{c}{-7.59} \\
          & \multicolumn{1}{c}{Eadro} & \multicolumn{1}{c}{49.35} & \multicolumn{1}{c}{31.39} & \multicolumn{1}{c}{-17.96} \\
\cmidrule{2-5}          & \multicolumn{1}{c}{Average} & \multicolumn{1}{c}{45.26} & \multicolumn{1}{c}{33.10} & \multicolumn{1}{c}{\cellcolor{pos}\textbf{-12.16}} \\
    \hline
    \multicolumn{1}{c}{\multirow{5}[4]{*}{HR@5}} & \multicolumn{1}{c}{RobustFlagger} & \multicolumn{1}{c}{54.84} & \multicolumn{1}{c}{30.18} & \multicolumn{1}{c}{-24.66} \\
          & \multicolumn{1}{c}{LogRCA} & \multicolumn{1}{c}{60.51} & \multicolumn{1}{c}{50.12} & \multicolumn{1}{c}{-10.39} \\
          & \multicolumn{1}{c}{AFAFormer} & \multicolumn{1}{c}{68.85} & \multicolumn{1}{c}{55.51} & \multicolumn{1}{c}{-13.34} \\
          & \multicolumn{1}{c}{Eadro} & \multicolumn{1}{c}{62.05} & \multicolumn{1}{c}{41.86} & \multicolumn{1}{c}{-20.19} \\
\cmidrule{2-5}          & \multicolumn{1}{c}{Average} & \multicolumn{1}{c}{61.56} & \multicolumn{1}{c}{44.42} & \multicolumn{1}{c}{\cellcolor{pos}\textbf{-17.14}} \\
    \hline
    \end{tabular}%
  \label{empirical_bias}%
  \vspace{-0.2cm}
\end{table}%

It is evident that existing methods do not effectively address diagnostic bias; the baseline methods show average reductions of 5.67\%, 12.16\%, and 17.14\% in HR@1, HR@3, and HR@5, respectively, indicating that diagnostic bias severely impacts fault diagnosis performance. This is because existing log-based fault diagnosis methods perform anomaly detection and root cause localization independently. If the detector makes an error, for instance by reporting a false positive, even if the localizer cannot identify the root cause, the localizer will still report a false positive root cause to the SRE due to the influence of the detector, resulting in a failed diagnosis. This provides motivation for us to collaborate the detector and localizer, as unifying the knowledge of the detector and localizer for a joint diagnosis would help reduce the noise transmission between them, thereby lowering diagnostic bias.

\begin{center}
    \begin{tcolorbox}[colback=gray!10,%gray background
        colframe=black,% black frame colour
        width=\linewidth,% Use 8cm total width,
        arc=1mm, auto outer arc,
        boxrule=0.5pt,
        top=2pt, % Adjust the top space
        bottom=2pt, % Adjust the bottom space
        left=2pt,
        right=2pt
        ]
        \textbf{Summary.} Existing log-based fault diagnosis methods conduct anomaly detection and root cause localization in a task-independent manner, which fails to adequately address the inevitable diagnostic bias in fault diagnosis, thereby severely impacting diagnostic performance.
    \end{tcolorbox}
\end{center}

\subsection{How effective are existing methods in addressing suboptimal diagnosis?} \label{sec:empirical_diag}
Log-based fault diagnosis is divided into two subtasks: anomaly detection and root cause localization, where the former detects whether a system fault has occurred, and the latter identifies the root cause of the fault. However, due to the differing diagnostic granularity of the two tasks, the diagnostic results for the same system fault may vary, and a system fault may not be simultaneously detected and localized. Specifically, based on the diagnostic results, faults can be classified into four categories: 
1. Detected and Localized Fault (\textbf{DLF}),
2. Detected but not Localized Fault (\textbf{DF}), 
3. Localized but not Detected Fault (\textbf{LF}) ,
4. Missing Fault (\textbf{MF}).

Among these, DLF represents the successful diagnosis, identified as optimal diagnosis, while MF indicates a complete failure in diagnosis, and DF and LF denote incomplete successful diagnosis, identified as suboptimal diagnosis. DF indicates that the fault can be effectively handled by the detector but not by the localizer, while LF is the opposite. Suboptimal diagnosis arises from the limited diagnostic capability of either the detector or the localizer, and by enhancing the diagnostic capabilities of the detector or localizer, suboptimal diagnosis can be transformed into optimal diagnosis.

\begin{figure}[htbp]
\centerline{
\includegraphics[width=0.8\linewidth]{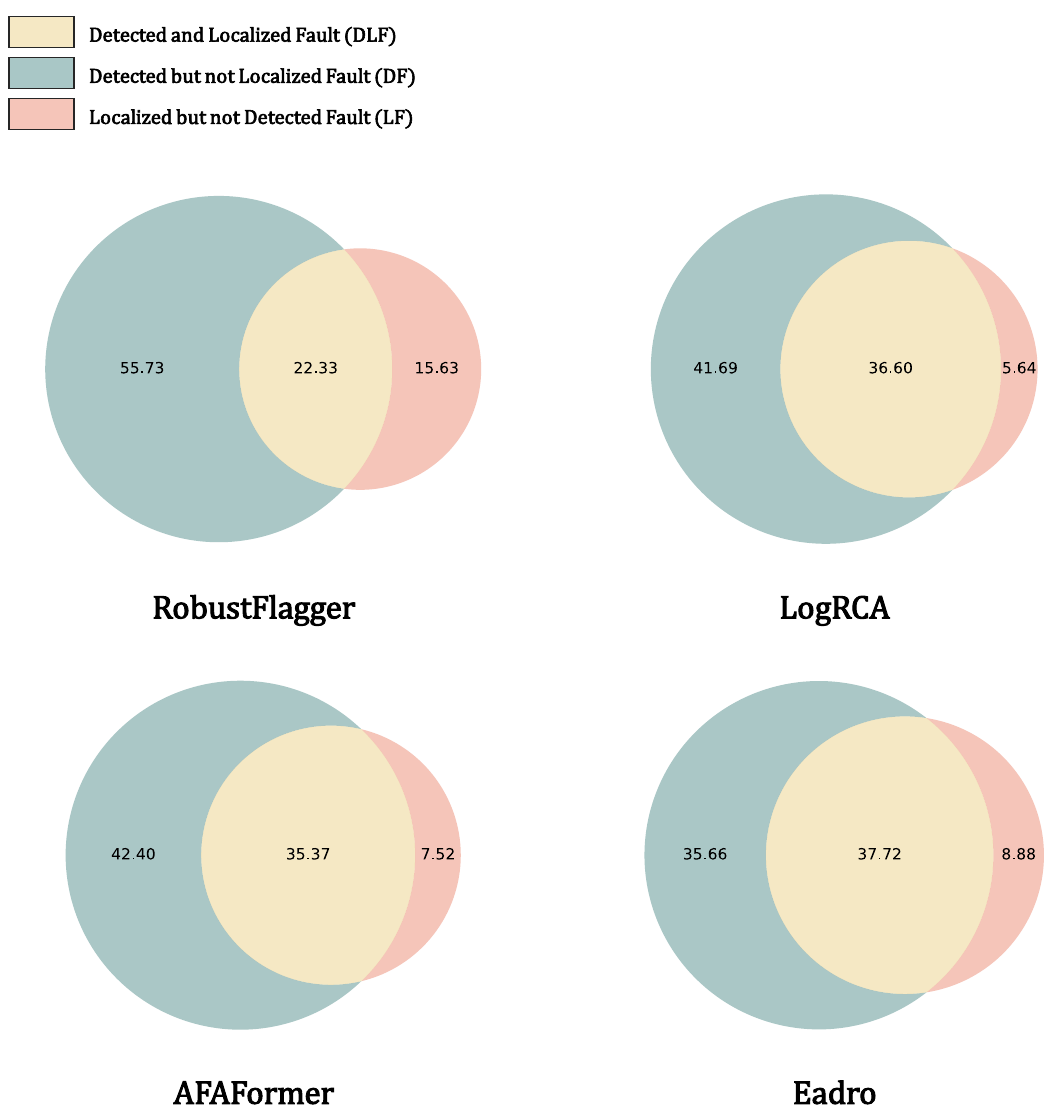}
}
\caption{The distribution of diagnostic results from existing methods in performing fault diagnosis on the Thunderbird dataset. The green portion represents Detected but not Localized Fault (DF), the red portion represents Localized but not Detected Fault (LF), and the yellow portion represents Detected and Localized Fault (DLF).}
\label{coll_thunder}
\vspace{-0.4cm}
\end{figure}

To study the effectiveness of existing methods in handling suboptimal diagnosis, we examined the distribution of suboptimal diagnoses DF and LF, as well as the optimal diagnosis DLF, generated by these methods when diagnosing the Thunderbird system, as shown in figure \ref{coll_thunder}. Please note that to eliminate the impact of inaccurate anomaly detection, we passed all system anomalies to the localizer.

Overall, existing methods produced a significant amount of suboptimal diagnoses, particularly of the DF type. This indicates that many faults are detected by the detector but fail during the root cause localization process. This is because existing methods perform anomaly detection and root cause localization independently, neglecting the collaborative relationship between the two diagnostic tasks, resulting in system faults not being simultaneously detected and localized, which leads to undesirable suboptimal diagnoses. This provides motivation for us to collaborate the detector and localizer, as leveraging the knowledge of the detector or localizer to enhance the diagnostic capabilities of the other could facilitate converting suboptimal diagnoses DF and LF into the optimal diagnosis DLF, thereby improving fault diagnosis performance.

\begin{center}
    \begin{tcolorbox}[colback=gray!10,%gray background
        colframe=black,% black frame colour
        width=\linewidth,% Use 8cm total width,
        arc=1mm, auto outer arc,
        boxrule=0.5pt,
        top=2pt, % Adjust the top space
        bottom=2pt, % Adjust the bottom space
        left=2pt,
        right=2pt
        ]
        \textbf{Summary.} Existing fault diagnosis methods independently perform anomaly detection and root cause localization, neglecting their collaborative relationship, which prevents simultaneous detection and localization of faults, resulting in a large number of undesirable suboptimal diagnoses.
    \end{tcolorbox}
\end{center}

\begin{figure*}[htbp]
\centerline{
\includegraphics[width=0.9\linewidth]{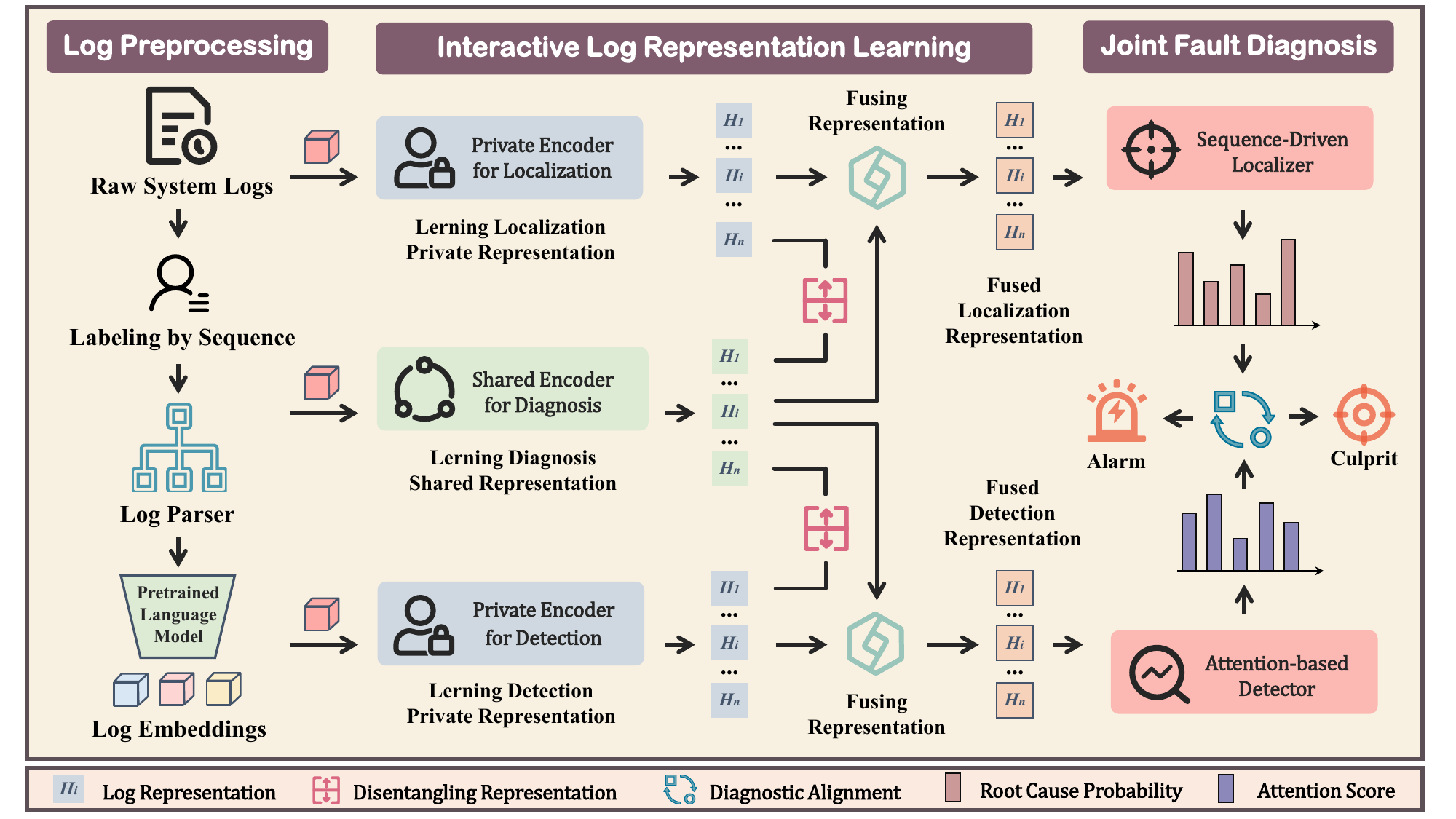}
}
\caption{The pipeline for \textbf{Chimera}. 
% Chimera pipeline comprises three key stages: Log Preprocessing, Interactive Log Representation Learning (ILRL), and Joint Fault Diagnosis. 
Firstly, the raw system logs are labeled and parsed into log event sequences, and corresponding log embeddings are extracted. Secondly, the ILRL module to learns shared and private representations for detection and localization interactively, and combines them into a log representation for fault diagnosis. Finally, the learned log representation is fed into the localizer and detector for joint fault diagnosis. Chimera bridges the gap between anomaly detection and root cause localization through their bidirectional interaction and knowledge transfer, achieving effective fault diagnosis.}
\label{pipeline}
\vspace{-0.4cm}
\end{figure*}

\section{Methodology}

\subsection{Overview}
In this paper, we present Chimera, an innovative end-to-end log-based fault diagnosis method designed to interactively achieve both anomaly detection and root cause localization within a unified framework. Log-based fault diagnosis is crucial for maintaining the reliability of industrial systems; however, existing methods construct models in a task-independent manner, which fails to provide precise and unified fault diagnosis. 
% To address this issue, we introduce Chimera, which primarily consists of three key processes: Log Preprocessing, Interactive Log Representation Learning, and Joint Fault Diagnosis. By leveraging the unique advantages they offer, Chimera can bridge the gap between anomaly detection and root cause localization in terms of data forms and diagnostic objectives, achieving more effective end-to-end fault diagnosis through their bidirectional interaction and knowledge transfer. Figure \ref{pipeline} illustrates the pipeline of Chimera.
To address this, we introduce Chimera, which can bridge the gap between anomaly detection and root cause localization, achieving more effective end-to-end fault diagnosis through their bidirectional interaction and knowledge transfer. Figure \ref{pipeline} illustrates the pipeline of Chimera.

Chimera first employs the advanced log parsing method, Drain \cite{he2017drain}, to process unstructured raw logs from software systems to obtain log events. Chimera follows existing work \cite{yang2021semi, zhang2024metalog} to extract semantic embeddings for each log event. 
% There is evidence \cite{zhang2019robust} suggesting that semantic embeddings provide richer information representation compared to index-based embedding methods. 
Next, Chimera performs the Interactive Log Representation Learning to learn interactive log representations from the log embeddings. This stage disentangles task-related private and shared representations and integrates them interactively, aiming to facilitate feature knowledge sharing between anomaly detection and root cause localization. Finally, the learned interactive log representations are fed into the Joint Fault Diagnosis stage for diagnosing system faults. 
Benefiting from the interactions of the diagnostic result, Chimera can leverage insights from the detector and localizer regarding anomalies of varying granularity, achieving precise fault diagnosis.

\subsection{Interactive Log Representation Learning}
Interactive Log Representation Learning aims to learn interactive log representations from log embeddings to facilitate feature knowledge sharing between the two tasks, bridging the gap in diagnostic objectives. In previous fault diagnosis work \cite{lee2023eadro, zhang2021onion, huo2023evlog, wittkopp2024logrca}, the features of the detector and localizer were either completely unrelated or fully shared. The former did not utilize the interaction information between tasks, while the latter overlooked the private information of each task. Numerous studies \cite{misra2016cross, ruder2019latent, lin2020shared} have demonstrated that excessive sharing of features is detrimental in multi-task interactions. How to achieve moderate feature interaction has become a significant challenge. To address this, Chimera has designed a private-shared feature interaction strategy, which involves disentangling the most necessary features to be shared from the log embeddings and integrating them with task-specific private features to achieve moderate interaction.

\subsubsection{\textbf{Shared-Private Log Representation Encoder}}
To achieve optimal interaction of task features, Chimera has designed the Shared-Private Log Representation Encoder (SPE), which aims to separately encode task-shared and task-private features from log embeddings. As shown in Figure \ref{pipeline}, the SPE consists of three independent encoders, of which two are private encoders, $\mathbf{E}_{p}^d$ and $\mathbf{E}_{p}^l$, responsible for encoding task-specific features, and one is a shared encoder, $\mathbf{E}_{s}$, responsible for encoding task-shared features. Specifically, given a log event embedding sequence of length $n$, $\mathbf{x} = [e_1, e_2, ..., e_n ]$, each encoder encodes specific representations:

\begin{equation}
    \begin{aligned}
\mathbf{V}_{p}^d &= \mathbf{E}_{p}^d (x) = [H_{p1}^d, H_{p2}^d, ..., H_{pn}^d ],  \\
\mathbf{V}_{p}^l &= \mathbf{E}_{p}^l (x) = [H_{p1}^l, H_{p2}^l, ..., H_{pn}^l ],  \\
\mathbf{V}_{s} &= \mathbf{E}_{s}(x) = [H_{s1}, H_{s2}, ..., H_{s2} ], 
    \end{aligned}
\end{equation}

Then, the SPE performs interactive fusion of the private representations and shared representations to obtain features for the anomaly detection and the root cause localization.
\begin{equation}
    \begin{aligned}
\mathbf{V}^d &= \mathbf{V}_p^d + \mathbf{V}_s  = [H_1^d, H_2^d, ..., H_n^d ],  \\
\mathbf{V}^l &= \mathbf{V}_p^l + \mathbf{V}_s  = [H_1^l, H_2^l, ..., H_n^l ], 
    \end{aligned}
\end{equation}
In log-based fault diagnosis, the log sequence consists of a series of log events that are continuously generated during system operation and are closely related within a short time frame. To effectively capture the temporal properties exhibited by log sequences, inspired by \cite{yang2021semi, zhang2024metalog}, Chimera has designed a Gated Recurrent Unit to implement the private and shared encoders $\mathbf{E}$ of the SPE. For each timestamp $t$, the encoder $\mathbf{E}$ maintains an update gate and a reset gate:
\begin{equation}
    \begin{aligned}
z_t &=\sigma\left(W_z\cdot[H_{t-1},e_t]\right),  \\
r_{t} &=\sigma\left(W_{r}\cdot[H_{t-1},e_{t}]\right),  \\
    \end{aligned}
\end{equation}
Here, $\sigma$ represents the logistic sigmoid function, $W$ denotes the network parameters, $H_{t-1}$ is the hidden state of the $t-1$th log event, and $e_t$ is the embedding of the $t$ log event in the input sequence. Subsequently, the encoder $\mathbf{E}$ computes the hidden state of the $t$h log event: 
\begin{equation}
    \begin{aligned}
\tilde{H}_{t} &=\operatorname{tanh}\left(W\cdot[r_{t}*H_{t-1},e_{t}]\right),  \\
H_t &=(1-z_t)*H_{t-1}+z_t*\tilde{H}_t, 
    \end{aligned}
\end{equation}
This design utilizes the update gate and reset gate to jointly determine the influence of past log events on future ones within a log sequence, allowing Chimera to capture the temporal information of the input log event sequence.

\subsubsection{\textbf{Disentanglement Interaction Loss}}
A significant challenge in merging the representations of the anomaly detection and the root cause localization is determining the appropriate shared representation. To address this, inspired by \cite{yue2022dare, nakano2023interaction, bousmalis2016domain}, Chimera proposes a disentanglement interaction loss, based on the key idea that the representations to be shared should differ as much as possible from the task-specific representations.

Specifically, if $\mathbf{V}_{p}^d$ is the private representation for the anomaly detection task encoded by the SPE, $\mathbf{V}_{p}^l$ is the private representation for the root cause localization task encoded by the SPE, and $\mathbf{V}_{s}$ is the shared representation between the two tasks, Chimera uses a soft subspace orthogonality constraint to define the loss:
\begin{equation}
\arg\min_\theta \mathcal{L}_{\text{Disentanglement}}(\theta) = \left\|\mathbf{V}_{p}^d \mathbf{V}_{s}^{\top} \right\|_{H}^{2} +\left\|\mathbf{V}_{p}^l \mathbf{V}_{s}^{\top} \right\|_{H}^{2}, 
\end{equation}

Here, $\left\|  \cdot \right\|_{H}^{2}$ represents the square of the Hilbert-Schmidt norm. This loss encourages the shared encoder and private encoder to encode representations as differently as possible, in order to disentangle the most necessary features to be shared.

\subsection{Joint Fault Diagnosis}
In this section, we introduce the Joint Fault Diagnosis stage, which utilizes the learned interactive log representations for end-to-end fault diagnosis. Chimera has designed an innovative localizer based on multi-instance learning, as well as a diagnostic result interaction strategy. Leveraging their advantages, Chimera is capable of training anomaly detection and root cause localization models using only log sequences, while aligning and interacting their diagnostic results to achieve accurate and unified end-to-end joint fault diagnosis.

\subsubsection{\textbf{Attention-based Detector}}
The log sequences in software systems are often highly complex, with noisy events significantly affecting the effectiveness of anomaly detection. To address these issues, Chimera has designed an attention-based detector for detecting anomalies from log sequences \cite{vaswani2017attention}. Specifically, for the input representation $\mathbf{V}^d = [H_1^d, H_2^d, ..., H_n^d ]$, the attention-based detector assigns a learnable weight to the hidden state: $\alpha^t = tanh(W^{\alpha}_t \cdot  H_t)$. 

Here, $W^{\alpha}_t$ is the learnable attention matrix. Subsequently, the detector weights the hidden states of the log events according to these weights, resulting in the representation of the log event sequence $\sum_{t = 0}^{t = n}\alpha^t \cdot H_t$. Finally, Chimera optimizes the detector by minimizing the cross-entropy loss.
\begin{equation}
\small
\arg\min_\theta \ \mathcal{L} _{\text{Detector}} (\theta) = \sum_{i=1}^{n}[-(y_{i}\log(\hat{y}_{i})+(1-y_{i})\log(1-\hat{y}_{i}))], 
\end{equation}

\begin{figure}[tbp]
\centerline{
\includegraphics[width=\linewidth]{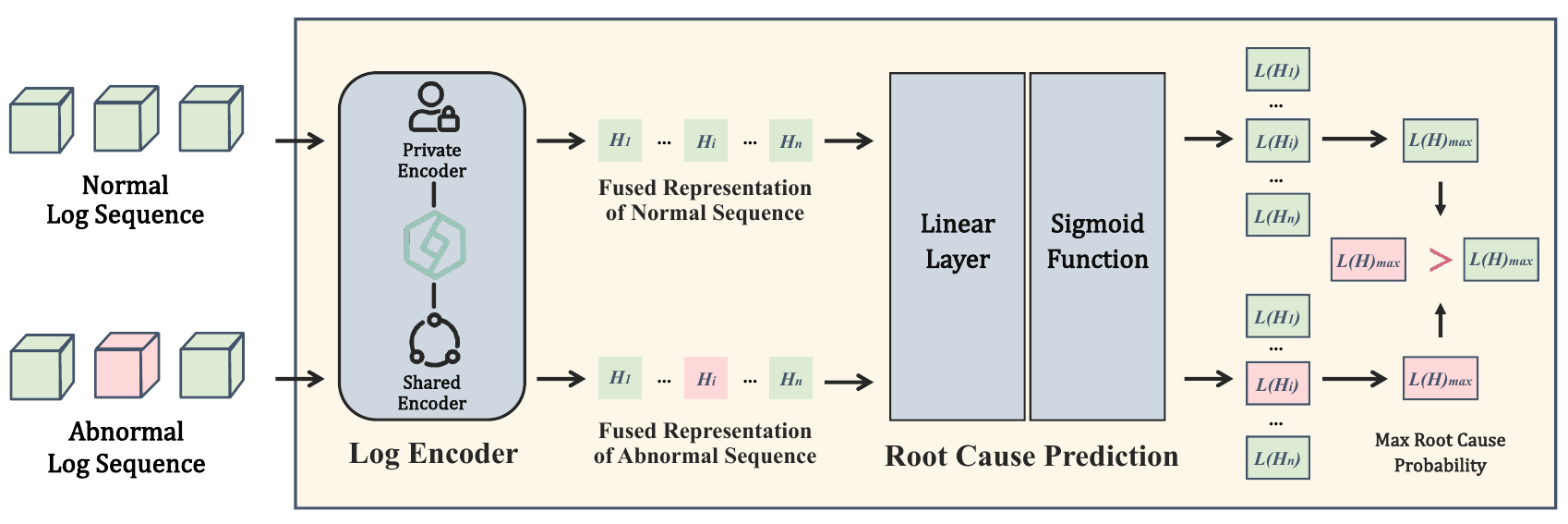}
}
\caption{The workflow of Sequence-driven Localizer. Given embeddings of log sequence, the network outputs a score for each log. The localizer is designed based on multi-instance learning, and locates the root cause log by comparing the scores of normal log sequences and anomalous log sequences.}
\label{localizer}
\vspace{-0.4cm}
\end{figure}

\subsubsection{\textbf{Sequence-driven Localizer}}
Existing methods require the simultaneous provision of anomalous log sequences and root cause logs to separately train the detector and the localizer. However, collecting a substantial amount of high-quality data of both types demands huge human effort, which is nearly impractical \cite{duan2023aclog, zhang2024metalog}. To address this challenge, Chimera has designed a sequence-driven localizer that trains only on anomalous log sequences, bridging the gap in data forms.

We first introduce the workflow of the sequence-driven localizer, as shown in \ref{localizer}. For the input log sequence representation $\mathbf{V}^l = [H_1^l, H_2^l, ..., H_n^l ]$, the localizer $L$ assigns a score $L(H)$ to each log event representation $H$, with a higher score indicating a greater likelihood of being a root cause log. 

Next, we introduce the training process of the sequence-driven localizer. The key idea of the localizer is to continually compare normal log sequences with anomalous log sequences to infer the root causes of the anomalies from the anomalous log sequences. Specifically, for the input normal log sequence representation $\mathbf{V}^l_n$ and the anomalous log sequence representation $\mathbf{V}^l_a$, Chimera has designed a ranking loss based on the principle of multi-instance learning \cite{jiang2023weakly, carbonneau2018multiple, perini2023learning, midlog}. 
\begin{equation}
    \begin{aligned}
\arg\min_\theta& \ \mathcal{L} _{\text{Localizer}}(\theta) =  \max (0, 1 + L(\mathbf{V}^l_n) - L(\mathbf{V}^l_a) ) \\
&=   \max (0, 1 + \max_{H^l \in \mathbf{V}^l_n} L(H^l) - \max_{H^l \in \mathbf{V}^l_a} L(H^l) ), 
    \end{aligned}
\end{equation}

This loss encourages the localizer to assign higher scores to log events in anomalous log sequences compared to those in normal log sequences, thereby inferring the root causes.

\subsubsection{\textbf{Cross-granularity Diagnostic Alignment}}
To further bridge the gap in diagnostic objectives between anomaly detection and root cause localization, and to promote knowledge transfer between the two tasks, Chimera has designed a cross-granularity diagnostic alignment module that facilitates interactive learning at diagnostic results. The fundamental insight is that the two subtasks have different diagnostic objectives for the faults, and bidirectional interaction supplements distinct granularities of diagnostic insights for each task.

Specifically, Chimera facilitates bidirectional interaction by constraining the two subtasks to have similar diagnostic results. This interaction strategy has been shown to effectively facilitate knowledge transfer between tasks \cite{chen2022hierarchical, su2023towards, bachman2019learning}. Given that the diagnostic objectives of the two tasks differ, making direct alignment of diagnostic results impossible, Chimera's strategy is to align the diagnostic results of the localizer with the attention scores of the detector, specifically by maximizing their mutual information. If $\mathbf{A} $ represents the distribution of the detector's attention scores and $\mathbf{R} $ denotes the distribution of the root causes identified by the localizer, Chimera's objective is: $\arg\max_{\theta}\ \mathcal{I}(\mathbf{A};\mathbf{R})$. 

To achieve this, an intuitive idea is to use the Kullback-Leibler (KL) divergence to measure the distance between $\mathbf{A}$ and $\mathbf{R}$, and then minimize this distance: $\arg\min_{\theta}\ KL(\mathbf{A}\|\mathbf{R})$. However, the KL divergence is asymmetric, and it can only facilitate unidirectional interaction. To this end, we adopt the Jensen-Shannon (JS) divergence to assess the similarity between $\mathbf{A}$ and $\mathbf{R}$, as its symmetric nature guarantees bidirectional interaction among the subtasks.

\begin{equation}
\small
JS(\mathbf{A}\|\mathbf{R})=\frac12KL\left(\mathbf{A}\|\frac{\mathbf{A}+\mathbf{R}}2\right)+\frac12KL\left(\mathbf{R}\|\frac{\mathbf{A}+\mathbf{R}}2\right), 
\end{equation}

\begin{equation}
\arg\min_{\theta} \ \mathcal{L} _{\text{Align}}(\theta) = \arg\min_{\theta}\ JS(\mathbf{A}\|\mathbf{R}), 
\end{equation}

\subsection{Training}
In this manner, Chimera can be deployed on real industrial software systems for end-to-end log-based fault diagnosis. Chimera performs end-to-end training by integrating the above-mentioned loss functions:
\begin{equation}
    \begin{aligned}
\arg\min_{\theta} \ \mathcal{L} (\theta) &= \lambda_1 \mathcal{L} _{\text{Detector}} (\theta) + \lambda_2 \mathcal{L} _{\text{Localizer}} (\theta)  \\
&+ \lambda_3 \mathcal{L} _{\text{Disentanglement}} (\theta)+ \lambda_4 \mathcal{L} _{\text{Align}} (\theta), 
    \end{aligned}
\end{equation}
 Here, $\lambda1,  \lambda2, \lambda3, \lambda4$ are hyperparameters used to balance the loss functions.

\begin{table*}[htbp]
  \centering
  \caption{Comparison results with baseline methods for root cause localization. All system anomalies are passed to the localizer.}
  \scalebox{0.85}{
    \begin{tabular}{rrrrrrrrrr}
    \hline
    \multicolumn{1}{c}{\textbf{Dataset}} & \multicolumn{1}{c}{\textbf{Model}} & \multicolumn{1}{c}{\textbf{HR@1}} & \multicolumn{1}{c}{\textbf{HR@3}} & \multicolumn{1}{c}{\textbf{HR@5}} & \multicolumn{1}{c}{\textbf{PR@3}} & \multicolumn{1}{c}{\textbf{PR@5}} & \multicolumn{1}{c}{\textbf{MAP@3}} & \multicolumn{1}{c}{\textbf{MAP@5}} & \multicolumn{1}{c}{\textbf{MRR}} \\
    \midrule
    \multicolumn{1}{c}{\multirow{5}[2]{*}{BGL}} & \multicolumn{1}{c}{LogFaultFlagger} & \multicolumn{1}{c}{40.64} & \multicolumn{1}{c}{67.18} & \multicolumn{1}{c}{73.43} & \multicolumn{1}{c}{44.20} & \multicolumn{1}{c}{46.74} & \multicolumn{1}{c}{42.42} & \multicolumn{1}{c}{43.94} & \multicolumn{1}{c}{56.53} \\
          & \multicolumn{1}{c}{LogRCA} & \multicolumn{1}{c}{42.89 ± 0.50} & \multicolumn{1}{c}{67.93 ± 1.55} & \multicolumn{1}{c}{78.38 ± 0.53} & \multicolumn{1}{c}{44.19 ± 0.25} & \multicolumn{1}{c}{47.10 ± 0.29} & \multicolumn{1}{c}{43.43 ± 0.26} & \multicolumn{1}{c}{44.56 ± 0.22} & \multicolumn{1}{c}{58.50 ± 0.45} \\
          & \multicolumn{1}{c}{AFAFormer} & \multicolumn{1}{c}{72.66 ± 3.11} & \multicolumn{1}{c}{88.98 ± 1.01} & \multicolumn{1}{c}{93.24 ± 0.62} & \multicolumn{1}{c}{77.81 ± 1.47} & \multicolumn{1}{c}{79.45 ± 1.15} & \multicolumn{1}{c}{74.68 ± 2.46} & \multicolumn{1}{c}{76.47 ± 1.57} & \multicolumn{1}{c}{80.48 ± 2.08} \\
          & \multicolumn{1}{c}{Eadro} & \multicolumn{1}{c}{45.34 ± 0.82} & \multicolumn{1}{c}{68.86 ± 0.81} & \multicolumn{1}{c}{79.82 ± 0.62} & \multicolumn{1}{c}{48.39 ± 0.63} & \multicolumn{1}{c}{51.91 ± 0.58} & \multicolumn{1}{c}{46.80 ± 0.62} & \multicolumn{1}{c}{48.50 ± 0.58} & \multicolumn{1}{c}{60.32 ± 0.68} \\
          & \multicolumn{1}{c}{Chimera} & \multicolumn{1}{c}{\cellcolor{pos}\textbf{89.73} ± \textbf{2.81}} & \multicolumn{1}{c}{\cellcolor{pos}\textbf{94.29} ± \textbf{1.19}} & \multicolumn{1}{c}{\cellcolor{pos}\textbf{95.76} ± \textbf{0.73}} & \multicolumn{1}{c}{\cellcolor{pos}\textbf{86.20} ± \textbf{2.44}} & \multicolumn{1}{c}{\cellcolor{pos}\textbf{81.28} ± \textbf{1.56}} & \multicolumn{1}{c}{\cellcolor{pos}\textbf{87.79} ± \textbf{2.21}} & \multicolumn{1}{c}{\cellcolor{pos}\textbf{85.61} ± \textbf{1.91}} & \multicolumn{1}{c}{\cellcolor{pos}\textbf{91.15} ± \textbf{1.77}} \\
    \midrule
    \multicolumn{1}{c}{\multirow{5}[2]{*}{Thunderbird}} & \multicolumn{1}{c}{LogFaultFlagger} & \multicolumn{1}{c}{19.83} & \multicolumn{1}{c}{39.21} & \multicolumn{1}{c}{54.84} & \multicolumn{1}{c}{30.35} & \multicolumn{1}{c}{42.39} & \multicolumn{1}{c}{24.96} & \multicolumn{1}{c}{30.73} & \multicolumn{1}{c}{36.49} \\
          & \multicolumn{1}{c}{LogRCA} & \multicolumn{1}{c}{21.00 ± 1.71} & \multicolumn{1}{c}{45.21 ± 0.95} & \multicolumn{1}{c}{60.51 ± 1.71} & \multicolumn{1}{c}{29.69 ± 1.33} & \multicolumn{1}{c}{40.99 ± 1.81} & \multicolumn{1}{c}{25.10 ± 1.48} & \multicolumn{1}{c}{30.27 ± 1.51} & \multicolumn{1}{c}{38.99 ± 1.14} \\
          & \multicolumn{1}{c}{AFAFormer} & \multicolumn{1}{c}{12.55 ± 6.59} & \multicolumn{1}{c}{47.27 ± 8.95} & \multicolumn{1}{c}{68.85 ± 8.41} & \multicolumn{1}{c}{40.84 ± 7.34} & \multicolumn{1}{c}{57.68 ± 6.44} & \multicolumn{1}{c}{27.32 ± 6.81} & \multicolumn{1}{c}{37.92 ± 6.68} & \multicolumn{1}{c}{35.92 ± 6.10} \\
          & \multicolumn{1}{c}{Eadro} & \multicolumn{1}{c}{28.41 ± 0.25} & \multicolumn{1}{c}{49.35 ± 0.24} & \multicolumn{1}{c}{62.05 ± 0.45} & \multicolumn{1}{c}{37.04 ± 0.23} & \multicolumn{1}{c}{46.82 ± 0.24} & \multicolumn{1}{c}{32.63 ± 0.16} & \multicolumn{1}{c}{37.30 ± 0.12} & \multicolumn{1}{c}{44.39 ± 0.16} \\
          & \multicolumn{1}{c}{Chimera} & \multicolumn{1}{c}{\cellcolor{pos}\textbf{68.57} ± \textbf{1.03}} & \multicolumn{1}{c}{\cellcolor{pos}\textbf{74.86} ± \textbf{0.72}} & \multicolumn{1}{c}{\cellcolor{pos}\textbf{76.57} ± \textbf{1.21}} & \multicolumn{1}{c}{\cellcolor{pos}\textbf{68.38} ± \textbf{2.16}} & \multicolumn{1}{c}{\cellcolor{pos}\textbf{69.03} ± \textbf{2.14}} & \multicolumn{1}{c}{\cellcolor{pos}\textbf{68.61} ± \textbf{1.61}} & \multicolumn{1}{c}{\cellcolor{pos}\textbf{68.59} ± \textbf{1.82}} & \multicolumn{1}{c}{\cellcolor{pos}\textbf{72.22} ± \textbf{0.79}} \\
    \midrule
    \multicolumn{1}{c}{\multirow{5}[2]{*}{System A}} & \multicolumn{1}{c}{LogFaultFlagger} & \multicolumn{1}{c}{56.59} & \multicolumn{1}{c}{67.90} & \multicolumn{1}{c}{74.75} & \multicolumn{1}{c}{58.69} & \multicolumn{1}{c}{61.82} & \multicolumn{1}{c}{57.59} & \multicolumn{1}{c}{58.92} & \multicolumn{1}{c}{65.48} \\
          & \multicolumn{1}{c}{LogRCA} & \multicolumn{1}{c}{56.42 ± 0.33} & \multicolumn{1}{c}{73.40 ± 1.00} & \multicolumn{1}{c}{81.36 ± 0.86} & \multicolumn{1}{c}{57.65 ± 0.23} & \multicolumn{1}{c}{60.35 ± 0.39} & \multicolumn{1}{c}{57.00 ± 0.20} & \multicolumn{1}{c}{58.05 ± 0.20} & \multicolumn{1}{c}{67.95 ± 0.20} \\
          & \multicolumn{1}{c}{AFAFormer} & \multicolumn{1}{c}{69.66 ± 1.39} & \multicolumn{1}{c}{79.15 ± 0.73} & \multicolumn{1}{c}{86.72 ± 0.53} & \multicolumn{1}{c}{73.01 ± 1.22} & \multicolumn{1}{c}{76.39 ± 1.42} & \multicolumn{1}{c}{71.18 ± 1.05} & \multicolumn{1}{c}{72.92 ± 1.08} & \multicolumn{1}{c}{79.24 ± 0.70} \\
          & \multicolumn{1}{c}{Eadro} & \multicolumn{1}{c}{59.42 ± 0.82} & \multicolumn{1}{c}{73.78 ± 0.26} & \multicolumn{1}{c}{81.82 ± 0.52} & \multicolumn{1}{c}{60.94 ± 0.40} & \multicolumn{1}{c}{64.10 ± 0.12} & \multicolumn{1}{c}{60.21 ± 0.53} & \multicolumn{1}{c}{61.43 ± 0.36} & \multicolumn{1}{c}{69.51 ± 0.42} \\
          & \multicolumn{1}{c}{Chimera} & \multicolumn{1}{c}{\cellcolor{pos}\textbf{75.28} ± \textbf{2.23}} & \multicolumn{1}{c}{\cellcolor{pos}\textbf{85.60} ± \textbf{1.43}} & \multicolumn{1}{c}{\cellcolor{pos}\textbf{88.55} ± \textbf{1.15}} & \multicolumn{1}{c}{\cellcolor{pos}\textbf{76.41} ± \textbf{1.22}} & \multicolumn{1}{c}{\cellcolor{pos}\textbf{77.07} ± \textbf{1.30}} & \multicolumn{1}{c}{\cellcolor{pos}\textbf{76.13} ± \textbf{1.39}} & \multicolumn{1}{c}{\cellcolor{pos}\textbf{76.39} ± \textbf{1.21}} & \multicolumn{1}{c}{\cellcolor{pos}\textbf{81.23} ± \textbf{2.79}} \\
    \hline
    \end{tabular}%
    }
  \label{rq1:rca}%
  \vspace{-0.4cm}
\end{table*}%

\section{Experimental Evaluation}
In this section, we evaluate our proposed method by addressing the following research questions: 

\begin{itemize}
    \item RQ1: How effective is Chimera in the anomaly detection and root cause localization?
    \item RQ2: How effective is Chimera in the log-based fault diagnosis?
    \item RQ3: How do different modules contribute to Chimera?
    \item RQ4: How effective is Chimera in addressing diagnostic bias?
    \item RQ5: How effective is Chimera in addressing suboptimal diagnosis?
\end{itemize}

\subsection{Experimental Setup}\label{sec:setup}
\subsubsection{Datasets}
We conducted comprehensive experiments on two widely used log-based anomaly detection public datasets: BGL, Thunderbird \cite{oliner2007supercomputers}, and an industrial dataset System A. The BGL dataset is derived from the operational logs of the Blue Gene/L supercomputing system, which contains 128K processors. The Thunderbird dataset contains over 200 million log messages, collected a supercomputer with 9,024 processors and 27,072 GB of memory. The System A dataset is sourced from an industrial cloud platform, which is built on a Kubernetes-based microservices architecture. 
Considering the huge scale of the Thunderbird dataset, we followed the settings of the previous study \cite{le2021log, le2022log} and selected the earliest 10 million log messages for experimentation.

\subsubsection{Compared Methods}
To better evaluate the effectiveness of Chimera, we compare it with advanced deep learning-based log anomaly detection methods and root cause localization methods. \textbf{NeuralLog} \cite{le2021log}, \textbf{SwissLog} \cite{li2022swisslog}, \textbf{LogRobust} \cite{zhang2019robust}, and \textbf{CNN} \cite{lu2018detecting} are the leading anomaly detection methods that utilize supervised learning, employing neural networks to map log sequences to vectors and then using classification-based approaches for anomaly detection. \textbf{LogFaultFlagger} \cite{amar2019mining} and \textbf{LogRCA} \cite{wittkopp2024logrca} are the advanced root cause localization methods that utilize statistical models or recurrent neural networks to mine root causes from historical log sequences. \textbf{Eadro} \cite{lee2023eadro} is a advanced end-to-end multimodal fault diagnosis system; to ensure fairness in the experiments, only log data is used for fault diagnosis in the evaluation. \textbf{AFAFormer} \cite{duan2023afalog} trains a Transformer network using labeled log sequences for anomaly detection and employs the network's multi-head attention scores for root cause localization.

\subsubsection{Evaluation Metrics}
To comprehensively assess the effectiveness of Chimera in log-based fault diagnosis, we employ a comprehensive set of metrics for the evaluation. For the anomaly detection task, we evaluate using Precision, Recall, and F1 Score. For the root cause localization task, we evaluate using Hit Rate at K (HR@k), Precision at K (PR@k), Mean Average Precision at K (MAP@k), and Mean Reciprocal Rank at K (MRR). We set $k=1,3,5$ for evaluation. When $k=1$, the meanings of metrics other than HR are equivalent to HR@1, so we only report results for $k=3$ and $k=5$.

\subsubsection{Implementation Details}
To ensure the reproducibility of the research, we present detailed implementation specifics. Following the setup of existing works \cite{meng2019loganomaly, le2022log}, we group the dataset using a sliding window of size 20, parse the log messages with Drain, and split the dataset into training, testing, and validation sets in a ratio of 6:3:1. The AdmaW optimizer is used for training optimization, with a learning rate set to 1e-3. The hyperparameters used to balance the joint training loss functions, $\lambda1,  \lambda2, \lambda3, \lambda4$, are set to 1, 2, 0.001, and 0.5, respectively. For all experiments, we conducted five runs and reported the mean and standard deviation. 
% The experiments for Chimera were conducted on a Linux server equipped with an Intel Xeon(R) Silver 4214R 2.4GHz processor and an RTX 3090 with 24GB GPU memory. 
% For the baseline methods used for comparison in the experiments, we utilized their publicly available implementations and settings. 
Code is available at \url{https://github.com/hemh02/Chimera}

\begin{table}[htbp]
  \centering
  \caption{Comparison results with baseline methods for anomaly detection.}
  \scalebox{0.9}{
    \begin{tabular}{rrrrr}
    \hline
    \multicolumn{1}{c}{\textbf{Dataset}} & \multicolumn{1}{c}{\textbf{Model}} & \multicolumn{1}{c}{\textbf{Precision}} & \multicolumn{1}{c}{\textbf{Recall}} & \multicolumn{1}{c}{\textbf{F1-Score}} \\
    \midrule
    \multicolumn{1}{c}{\multirow{5}[2]{*}{BGL}} & \multicolumn{1}{c}{CNN} & \multicolumn{1}{c}{89.31 ± 0.76} & \multicolumn{1}{c}{70.38 ± 8.14} & \multicolumn{1}{c}{78.44 ± 5.09} \\
          & \multicolumn{1}{c}{LogRobust} & \multicolumn{1}{c}{90.31 ± 1.97} & \multicolumn{1}{c}{92.73 ± 1.31} & \multicolumn{1}{c}{91.47 ± 0.47} \\
          & \multicolumn{1}{c}{NeuralLog} & \multicolumn{1}{c}{86.97 ± 3.62} & \multicolumn{1}{c}{75.54 ± 2.57} & \multicolumn{1}{c}{80.82 ± 2.64} \\
          & \multicolumn{1}{c}{SwissLog} & \multicolumn{1}{c}{99.31 ± 0.26} & \multicolumn{1}{c}{84.43 ± 1.65} & \multicolumn{1}{c}{91.26 ± 0.93} \\
          & \multicolumn{1}{c}{Chimera} & \multicolumn{1}{c}{91.22 ± 1.90} & \multicolumn{1}{c}{97.37 ± 0.36} & \multicolumn{1}{c}{\cellcolor{pos}\textbf{94.19} ± \textbf{1.05}} \\ % 调整了这一行
    \midrule
    \multicolumn{1}{c}{\multirow{5}[2]{*}{Thunderbird}} & \multicolumn{1}{c}{CNN} & \multicolumn{1}{c}{94.38 ± 1.62} & \multicolumn{1}{c}{66.03 ± 1.91} & \multicolumn{1}{c}{77.67 ± 1.28} \\
          & \multicolumn{1}{c}{LogRobust} & \multicolumn{1}{c}{91.36 ± 5.80} & \multicolumn{1}{c}{78.13 ± 2.65} & \multicolumn{1}{c}{84.13 ± 3.17} \\
          & \multicolumn{1}{c}{NeuralLog} & \multicolumn{1}{c}{76.34 ± 4.40} & \multicolumn{1}{c}{74.46 ± 3.26} & \multicolumn{1}{c}{75.20 ± 1.17} \\
          & \multicolumn{1}{c}{SwissLog} & \multicolumn{1}{c}{31.74 ± 3.13} & \multicolumn{1}{c}{48.13 ± 8.85} & \multicolumn{1}{c}{38.16 ± 5.09} \\
          & \multicolumn{1}{c}{Chimera} & \multicolumn{1}{c}{98.08 ± 0.75} & \multicolumn{1}{c}{82.82 ± 2.18} & \multicolumn{1}{c}{\cellcolor{pos}\textbf{89.80} ± \textbf{1.59}} \\ % 调整了这一行
    \midrule
    \multicolumn{1}{c}{\multirow{5}[2]{*}{System A}} & \multicolumn{1}{c}{CNN} & \multicolumn{1}{c}{78.34 ± 2.79} & \multicolumn{1}{c}{68.87 ± 0.66} & \multicolumn{1}{c}{73.29 ± 1.55} \\
          & \multicolumn{1}{c}{LogRobust} & \multicolumn{1}{c}{82.35 ± 2.55} & \multicolumn{1}{c}{86.42 ± 1.82} & \multicolumn{1}{c}{84.28 ± 0.54} \\
          & \multicolumn{1}{c}{NeuralLog} & \multicolumn{1}{c}{54.40 ± 2.90} & \multicolumn{1}{c}{74.60 ± 2.18} & \multicolumn{1}{c}{66.04 ± 1.11} \\
          & \multicolumn{1}{c}{SwissLog} & \multicolumn{1}{c}{88.17 ± 4.42} & \multicolumn{1}{c}{78.35 ± 3.39} & \multicolumn{1}{c}{82.80 ± 1.19} \\
          & \multicolumn{1}{c}{Chimera} & \multicolumn{1}{c}{82.75 ± 1.02} & \multicolumn{1}{c}{93.30 ± 1.57} & \multicolumn{1}{c}{\cellcolor{pos}\textbf{87.70} ± \textbf{0.58}} \\ % 调整了这一行
    \hline
    \end{tabular}%
    }
  \label{rq1:ad}%
  \vspace{-0.4cm}
\end{table}%

\begin{table*}[htbp]
  \centering
  \caption{Comparison results with baseline methods for fault diagnosis in root cause localization task.}
  \scalebox{0.78}{
    \begin{tabular}{rrrrrrrrrrr} % 调整了列数，原始为rrrrrrrrrrrrr，但实际内容只有11列
    \hline
    \multicolumn{1}{c}{\textbf{Dataset}} & \multicolumn{1}{c}{\textbf{Deployment}} & \multicolumn{1}{c}{\textbf{Model}} & \multicolumn{1}{c}{\textbf{HR@1}} & \multicolumn{1}{c}{\textbf{HR@3}} & \multicolumn{1}{c}{\textbf{HR@5}} & \multicolumn{1}{c}{\textbf{PR@3}} & \multicolumn{1}{c}{\textbf{PR@5}} & \multicolumn{1}{c}{\textbf{MAP@3}} & \multicolumn{1}{c}{\textbf{MAP@5}} & \multicolumn{1}{c}{\textbf{MRR}} \\
    \hline
    \multicolumn{1}{c}{\multirow{5}[6]{*}{BGL}} & \multicolumn{1}{p{8em}}{Task-Independent} & \multicolumn{1}{c}{RobustFlagger} & \multicolumn{1}{c}{40.42 ± 0.18} & \multicolumn{1}{c}{65.53 ± 0.21} & \multicolumn{1}{c}{71.68 ± 0.38} & \multicolumn{1}{c}{42.49 ± 0.24} & \multicolumn{1}{c}{45.13 ± 0.34} & \multicolumn{1}{c}{41.12 ± 0.21} & \multicolumn{1}{c}{42.45 ± 0.25} & \multicolumn{1}{c}{55.90 ± 0.38} \\
          & \multicolumn{1}{c}{(Pipeline)} & \multicolumn{1}{c}{LogRCA} & \multicolumn{1}{c}{39.53 ± 0.84} & \multicolumn{1}{c}{65.55 ± 2.05} & \multicolumn{1}{c}{73.98 ± 1.72} & \multicolumn{1}{c}{42.08 ± 0.64} & \multicolumn{1}{c}{44.72 ± 0.66} & \multicolumn{1}{c}{40.98 ± 0.78} & \multicolumn{1}{c}{42.16 ± 0.72} & \multicolumn{1}{c}{55.11 ± 1.30} \\
\cmidrule{2-11}          & \multicolumn{1}{p{8em}}{Task-Independent} & \multicolumn{1}{c}{AFAFormer} & \multicolumn{1}{c}{70.64 ± 3.23} & \multicolumn{1}{c}{85.27 ± 1.15} & \multicolumn{1}{c}{88.14 ± 1.28} & \multicolumn{1}{c}{78.65 ± 1.71} & \multicolumn{1}{c}{79.55 ± 1.15} & \multicolumn{1}{c}{76.03 ± 2.42} & \multicolumn{1}{c}{77.39 ± 1.76} & \multicolumn{1}{c}{79.38 ± 1.78} \\
          & \multicolumn{1}{c}{(End-to-End)} & \multicolumn{1}{c}{Eadro} & \multicolumn{1}{c}{38.88 ± 0.97} & \multicolumn{1}{c}{61.81 ± 1.32} & \multicolumn{1}{c}{69.71 ± 1.94} & \multicolumn{1}{c}{40.04 ± 0.68} & \multicolumn{1}{c}{41.22 ± 0.80} & \multicolumn{1}{c}{39.53 ± 0.77} & \multicolumn{1}{c}{40.04 ± 0.73} & \multicolumn{1}{c}{51.80 ± 1.15} \\
\cmidrule{2-11}          & \multicolumn{1}{c}{Task-Interactive} & \multicolumn{1}{c}{Chimera} & \multicolumn{1}{c}{\cellcolor{pos}\textbf{89.73} ± \textbf{2.81}} & \multicolumn{1}{c}{\cellcolor{pos}\textbf{94.29} ± \textbf{1.19}} & \multicolumn{1}{c}{\cellcolor{pos}\textbf{95.76} ± \textbf{0.73}} & \multicolumn{1}{c}{\cellcolor{pos}\textbf{86.20} ± \textbf{2.44}} & \multicolumn{1}{c}{\cellcolor{pos}\textbf{81.28} ± \textbf{1.56}} & \multicolumn{1}{c}{\cellcolor{pos}\textbf{87.79} ± \textbf{2.21}} & \multicolumn{1}{c}{\cellcolor{pos}\textbf{85.61} ± \textbf{1.91}} & \multicolumn{1}{c}{\cellcolor{pos}\textbf{91.15} ± \textbf{1.77}} \\
    \hline
    \multicolumn{1}{c}{\multirow{5}[6]{*}{Thunderbird}} & \multicolumn{1}{p{8em}}{Task-Independent} & \multicolumn{1}{c}{RobustFlagger} & \multicolumn{1}{c}{13.78 ± 0.08} & \multicolumn{1}{c}{22.81 ± 0.12} & \multicolumn{1}{c}{30.18 ± 0.17} & \multicolumn{1}{c}{15.23 ± 0.07} & \multicolumn{1}{c}{19.06 ± 0.13} & \multicolumn{1}{c}{14.36 ± 0.06} & \multicolumn{1}{c}{15.72 ± 0.08} & \multicolumn{1}{c}{29.77 ± 0.34} \\
          & \multicolumn{1}{c}{(Pipeline)} & \multicolumn{1}{c}{LogRCA} & \multicolumn{1}{c}{19.05 ± 0.95} & \multicolumn{1}{c}{38.50 ± 1.73} & \multicolumn{1}{c}{50.12 ± 2.44} & \multicolumn{1}{c}{24.78 ± 1.59} & \multicolumn{1}{c}{32.30 ± 2.68} & \multicolumn{1}{c}{21.45 ± 0.97} & \multicolumn{1}{c}{25.01 ± 1.56} & \multicolumn{1}{c}{33.04 ± 1.32} \\
\cmidrule{2-11}          & \multicolumn{1}{p{8em}}{Task-Independent} & \multicolumn{1}{c}{AFAFormer} & \multicolumn{1}{c}{11.68 ± 5.30} & \multicolumn{1}{c}{39.68 ± 8.91} & \multicolumn{1}{c}{55.51 ± 6.17} & \multicolumn{1}{c}{32.03 ± 6.25} & \multicolumn{1}{c}{44.81 ± 5.03} & \multicolumn{1}{c}{22.06 ± 5.99} & \multicolumn{1}{c}{30.04 ± 5.66} & \multicolumn{1}{c}{29.79 ± 5.71} \\
          & \multicolumn{1}{c}{(End-to-End)} & \multicolumn{1}{c}{Eadro} & \multicolumn{1}{c}{14.59 ± 1.27} & \multicolumn{1}{c}{31.39 ± 1.16} & \multicolumn{1}{c}{41.86 ± 1.31} & \multicolumn{1}{c}{16.88 ± 0.76} & \multicolumn{1}{c}{21.71 ± 0.90} & \multicolumn{1}{c}{15.69 ± 0.97} & \multicolumn{1}{c}{17.57 ± 0.87} & \multicolumn{1}{c}{27.50 ± 1.14} \\
\cmidrule{2-11}          & \multicolumn{1}{c}{Task-Interactive} & \multicolumn{1}{c}{Chimera} & \multicolumn{1}{c}{\cellcolor{pos}\textbf{68.57} ± \textbf{1.03}} & \multicolumn{1}{c}{\cellcolor{pos}\textbf{74.86} ± \textbf{0.72}} & \multicolumn{1}{c}{\cellcolor{pos}\textbf{76.57} ± \textbf{1.21}} & \multicolumn{1}{c}{\cellcolor{pos}\textbf{68.38} ± \textbf{2.16}} & \multicolumn{1}{c}{\cellcolor{pos}\textbf{69.03} ± \textbf{2.14}} & \multicolumn{1}{c}{\cellcolor{pos}\textbf{68.61} ± \textbf{1.61}} & \multicolumn{1}{c}{\cellcolor{pos}\textbf{68.59} ± \textbf{1.82}} & \multicolumn{1}{c}{\cellcolor{pos}\textbf{72.22} ± \textbf{0.79}} \\
    \hline
    \multicolumn{1}{c}{\multirow{5}[6]{*}{System A}} & \multicolumn{1}{p{8em}}{Task-Independent} & \multicolumn{1}{c}{RobustFlagger} & \multicolumn{1}{c}{54.97 ± 0.37} & \multicolumn{1}{c}{64.42 ± 0.69} & \multicolumn{1}{c}{70.00 ± 0.86} & \multicolumn{1}{c}{55.00 ± 0.43} & \multicolumn{1}{c}{56.93 ± 0.50} & \multicolumn{1}{c}{54.86 ± 0.40} & \multicolumn{1}{c}{55.46 ± 0.43} & \multicolumn{1}{c}{63.92 ± 0.53} \\
          & \multicolumn{1}{c}{(Pipeline)} & \multicolumn{1}{c}{LogRCA} & \multicolumn{1}{c}{54.33 ± 0.95} & \multicolumn{1}{c}{68.43 ± 2.27} & \multicolumn{1}{c}{75.42 ± 0.95} & \multicolumn{1}{c}{55.40 ± 0.46} & \multicolumn{1}{c}{57.42 ± 0.47} & \multicolumn{1}{c}{54.87 ± 0.67} & \multicolumn{1}{c}{55.68 ± 0.58} & \multicolumn{1}{c}{63.62 ± 0.70} \\
\cmidrule{2-11}          & \multicolumn{1}{p{8em}}{Task-Independent} & \multicolumn{1}{c}{AFAFormer} & \multicolumn{1}{c}{63.96 ± 2.43} & \multicolumn{1}{c}{77.78 ± 3.96} & \multicolumn{1}{c}{83.38 ± 3.67} & \multicolumn{1}{c}{66.18 ± 2.99} & \multicolumn{1}{c}{68.63 ± 3.46} & \multicolumn{1}{c}{64.96 ± 2.67} & \multicolumn{1}{c}{66.32 ± 2.95} & \multicolumn{1}{c}{71.82 ± 2.82} \\
          & \multicolumn{1}{c}{(End-to-End)} & \multicolumn{1}{c}{Eadro} & \multicolumn{1}{c}{52.36 ± 1.12} & \multicolumn{1}{c}{63.48 ± 1.57} & \multicolumn{1}{c}{68.31 ± 2.07} & \multicolumn{1}{c}{52.25 ± 1.03} & \multicolumn{1}{c}{53.90 ± 1.07} & \multicolumn{1}{c}{52.20 ± 1.01} & \multicolumn{1}{c}{52.69 ± 1.03} & \multicolumn{1}{c}{59.35 ± 1.41} \\
\cmidrule{2-11}          & \multicolumn{1}{c}{Task-Interactive} & \multicolumn{1}{c}{Chimera} & \multicolumn{1}{c}{\cellcolor{pos}\textbf{75.28} ± \textbf{2.23}} & \multicolumn{1}{c}{\cellcolor{pos}\textbf{85.60} ± \textbf{1.43}} & \multicolumn{1}{c}{\cellcolor{pos}\textbf{88.55} ± \textbf{1.15}} & \multicolumn{1}{c}{\cellcolor{pos}\textbf{76.41} ± \textbf{1.22}} & \multicolumn{1}{c}{\cellcolor{pos}\textbf{77.07} ± \textbf{1.30}} & \multicolumn{1}{c}{\cellcolor{pos}\textbf{76.13} ± \textbf{1.39}} & \multicolumn{1}{c}{\cellcolor{pos}\textbf{76.39} ± \textbf{1.21}} & \multicolumn{1}{c}{\cellcolor{pos}\textbf{81.23} ± \textbf{2.79}} \\
    \hline
    \end{tabular}%
    }
  \label{rq2:rca}%
  \vspace{-0.4cm}
\end{table*}%

\subsection{RQ1: How effective is Chimera in the anomaly detection and root cause localization?}
To evaluate the effectiveness of Chimera in log-based fault diagnosis, we evaluated Chimera separately on anomaly detection and root cause localization. Specifically, we compared Chimera with expert models for these two tasks, rather than complete fault diagnosis methods, to comprehensively evaluate Chimera's capabilities in anomaly detection and root cause localization. For the baseline setup in the root cause localization, we passed all system anomalies to the localizer to eliminate the impact of inaccurate anomaly detection. Tables \ref{rq1:rca} and \ref{rq1:ad} present the comparison with the baselines.

Overall, Chimera achieved best performance even when compared to expert models addressing anomaly detection and root cause localization. For the anomaly detection, the results obtained by Chimera exceeded those of the best supervised methods by 2.72\%, 5.67\%, and 3.42\% in F1 scores, respectively. For the root cause localization, Chimera demonstrated even more significant advantages. Compared to the advanced methods, it averaged 21.62\%, 12.42\%, and 4.02\% higher in HR@k; 13.11\%, 4.62\% higher in PR@k; 18.01\%, 14.43\% higher in MAP@k; and 13.49\% higher in MRR. This significant leap indicates that the bidirectional interaction and knowledge transfer between anomaly detection and root cause localization facilitate the effective utilization of diagnostic information, leading to more efficient fault diagnosis. 

% Additionally, in the evaluation of the root cause localization task, we observed that the baseline models performed disastrously on the PR@k metric. This is because existing root cause localization methods focus solely on identifying primary causes, neglecting sequence-level analysis of root causes, which leads to poor-quality root cause candidate lists \cite{zheng2024lemma}. 
% In contrast, Chimera supplements root cause localization with sequence-level diagnostic information through the bidirectional interaction of anomaly detection and root cause localization, resulting in more reliable root cause candidate lists. This evidence demonstrates the strong effectiveness of Chimera in fault diagnosis tasks.

\begin{table}[htbp]
  \centering
  \caption{Comparison results with baseline methods for fault diagnosis in anomaly detection task.}
    \scalebox{0.85}{
    \begin{tabular}{rrrrrr}
    \hline
    \multicolumn{1}{c}{\textbf{Dataset}} & \multicolumn{1}{c}{\textbf{Model}} & \multicolumn{1}{c}{\textbf{Precision}} & \multicolumn{1}{c}{\textbf{Recall}} & \multicolumn{1}{c}{\textbf{F1-Score}} \\
    \hline
    \multicolumn{1}{c}{\multirow{5}[6]{*}{BGL}} & \multicolumn{1}{c}{RobustFlagger} & \multicolumn{1}{c}{87.98 ± 5.89} & \multicolumn{1}{c}{92.83 ± 1.37} & \multicolumn{1}{c}{90.20 ± 2.64} \\
          & \multicolumn{1}{c}{LogRCA} & \multicolumn{1}{c}{84.93 ± 2.82} & \multicolumn{1}{c}{93.91 ± 0.86} & \multicolumn{1}{c}{89.16 ± 1.25} \\
\cmidrule{2-5}          & \multicolumn{1}{c}{AFAFormer} & \multicolumn{1}{c}{87.51 ± 3.76} & \multicolumn{1}{c}{92.59 ± 1.30} & \multicolumn{1}{c}{89.95 ± 2.35} \\
          & \multicolumn{1}{c}{Eadro} & \multicolumn{1}{c}{91.10 ± 0.97} & \multicolumn{1}{c}{77.78 ± 2.05} & \multicolumn{1}{c}{83.91 ± 1.51} \\
\cmidrule{2-5}          & \multicolumn{1}{c}{Chimera} & \multicolumn{1}{c}{91.22 ± 1.90} & \multicolumn{1}{c}{97.37 ± 0.36} & \multicolumn{1}{c}{\cellcolor{pos}\textbf{94.19} ± \textbf{1.05}} \\
    \hline
    \multicolumn{1}{c}{\multirow{5}[6]{*}{Thunderbird}} & \multicolumn{1}{c}{RobustFlagger} & \multicolumn{1}{c}{92.34 ± 4.30} & \multicolumn{1}{c}{78.62 ± 1.72} & \multicolumn{1}{c}{84.83 ± 1.06} \\
          & \multicolumn{1}{c}{LogRCA} & \multicolumn{1}{c}{96.33 ± 1.60} & \multicolumn{1}{c}{76.96 ± 3.85} & \multicolumn{1}{c}{85.49 ± 2.35} \\
\cmidrule{2-5}          & \multicolumn{1}{c}{AFAFormer} & \multicolumn{1}{c}{95.87 ± 1.10} & \multicolumn{1}{c}{76.10 ± 3.31} & \multicolumn{1}{c}{84.80 ± 1.92} \\
          & \multicolumn{1}{c}{Eadro} & \multicolumn{1}{c}{93.79 ± 0.25} & \multicolumn{1}{c}{73.76 ± 0.39} & \multicolumn{1}{c}{82.58 ± 0.31} \\
\cmidrule{2-5}          & \multicolumn{1}{c}{Chimera} & \multicolumn{1}{c}{98.08 ± 0.75} & \multicolumn{1}{c}{82.82 ± 2.18} & \multicolumn{1}{c}{\cellcolor{pos}\textbf{89.80} ± \textbf{1.59}} \\
    \hline
    \multicolumn{1}{c}{\multirow{5}[6]{*}{System A}} & \multicolumn{1}{c}{RobustFlagger} & \multicolumn{1}{c}{84.69 ± 2.06} & \multicolumn{1}{c}{84.95 ± 1.62} & \multicolumn{1}{c}{84.78 ± 0.53} \\
          & \multicolumn{1}{c}{LogRCA} & \multicolumn{1}{c}{78.70 ± 3.78} & \multicolumn{1}{c}{88.63 ± 1.87} & \multicolumn{1}{c}{83.27 ± 1.25} \\
\cmidrule{2-5}          & \multicolumn{1}{c}{AFAFormer} & \multicolumn{1}{c}{77.41 ± 2.89} & \multicolumn{1}{c}{89.66 ± 0.99} & \multicolumn{1}{c}{83.04 ± 1.43} \\
          & \multicolumn{1}{c}{Eadro} & \multicolumn{1}{c}{79.54 ± 1.97} & \multicolumn{1}{c}{77.21 ± 2.94} & \multicolumn{1}{c}{78.35 ± 2.44} \\
\cmidrule{2-5}          & \multicolumn{1}{c}{Chimera} & \multicolumn{1}{c}{82.75 ± 1.02} & \multicolumn{1}{c}{93.30 ± 1.57} & \multicolumn{1}{c}{\cellcolor{pos}\textbf{87.70} ± \textbf{0.58}} \\
    \hline
    \end{tabular}%
    }
  \label{rq2:ad}%
  \vspace{-0.4cm}
\end{table}%

\begin{table*}[htbp]
  \centering
  \caption{Ablation results on the root cause localization task.}
  \scalebox{0.85}{
    \begin{tabular}{rrrrrrrrrrr}
    \hline
    \multicolumn{1}{c}{\textbf{Dataset}} & \multicolumn{1}{c}{\textbf{ILRL}} & \multicolumn{1}{c|}{\textbf{CDA}} & \multicolumn{1}{c}{\textbf{HR@1}} & \multicolumn{1}{c}{\textbf{HR@3}} & \multicolumn{1}{c}{\textbf{HR@5}} & \multicolumn{1}{c}{\textbf{PR@3}} & \multicolumn{1}{c}{\textbf{PR@5}} & \multicolumn{1}{c}{\textbf{MAP@3}} & \multicolumn{1}{c}{\textbf{MAP@5}} & \multicolumn{1}{c}{\textbf{MRR}} \\
    \midrule
    \multicolumn{1}{c}{\multirow{4}[2]{*}{BGL}} & \multicolumn{1}{c}{\XSolidBrush} & \multicolumn{1}{c|}{\XSolidBrush} & \multicolumn{1}{c}{81.91 ± 3.79} & \multicolumn{1}{c}{87.91 ± 2.36} & \multicolumn{1}{c}{90.77 ± 1.50} & \multicolumn{1}{c}{78.45 ± 4.79} & \multicolumn{1}{c}{73.94 ± 4.22} & \multicolumn{1}{c}{80.29 ± 4.27} & \multicolumn{1}{c}{78.11 ± 4.24} & \multicolumn{1}{c}{85.92 ± 2.81} \\
          & \multicolumn{1}{c}{\XSolidBrush} & \multicolumn{1}{c|}{\Checkmark} & \multicolumn{1}{c}{82.34 ± 4.82} & \multicolumn{1}{c}{88.38 ± 2.11} & \multicolumn{1}{c}{91.17 ± 1.77} & \multicolumn{1}{c}{78.59 ± 4.92} & \multicolumn{1}{c}{74.57 ± 4.90} & \multicolumn{1}{c}{80.49 ± 5.00} & \multicolumn{1}{c}{78.53 ± 4.82} & \multicolumn{1}{c}{86.25 ± 3.33} \\
          & \multicolumn{1}{c}{\Checkmark} & \multicolumn{1}{c|}{\XSolidBrush} & \multicolumn{1}{c}{87.98 ± 2.39} & \multicolumn{1}{c}{93.12 ± 1.28} & \multicolumn{1}{c}{94.72 ± 0.91} & \multicolumn{1}{c}{83.91 ± 2.72} & \multicolumn{1}{c}{77.74 ± 3.17} & \multicolumn{1}{c}{86.38 ± 2.57} & \multicolumn{1}{c}{83.47 ± 2.75} & \multicolumn{1}{c}{91.08 ± 1.70} \\
          & \multicolumn{1}{c}{\Checkmark} & \multicolumn{1}{c|}{\Checkmark} & \multicolumn{1}{c}{\cellcolor{pos}\textbf{89.73} ± \textbf{2.81}} & \multicolumn{1}{c}{\cellcolor{pos}\textbf{94.29} ± \textbf{1.19}} & \multicolumn{1}{c}{\cellcolor{pos}\textbf{95.76} ± \textbf{0.73}} & \multicolumn{1}{c}{\cellcolor{pos}\textbf{86.20} ± \textbf{2.44}} & \multicolumn{1}{c}{\cellcolor{pos}\textbf{81.28} ± \textbf{1.56}} & \multicolumn{1}{c}{\cellcolor{pos}\textbf{87.79} ± \textbf{2.21}} & \multicolumn{1}{c}{\cellcolor{pos}\textbf{85.61} ± \textbf{1.91}} & \multicolumn{1}{c}{\cellcolor{pos}\textbf{91.15} ± \textbf{1.77}} \\
    \midrule
    \multicolumn{1}{c}{\multirow{4}[2]{*}{Thunderbird}} & \multicolumn{1}{c}{\XSolidBrush} & \multicolumn{1}{c|}{\XSolidBrush} & \multicolumn{1}{c}{55.86 ± 7.21} & \multicolumn{1}{c}{69.15 ± 5.10} & \multicolumn{1}{c}{73.01 ± 3.42} & \multicolumn{1}{c}{57.09 ± 6.37} & \multicolumn{1}{c}{60.49 ± 4.93} & \multicolumn{1}{c}{56.26 ± 6.77} & \multicolumn{1}{c}{57.60 ± 6.12} & \multicolumn{1}{c}{63.19 ± 5.42} \\
          & \multicolumn{1}{c}{\XSolidBrush} & \multicolumn{1}{c|}{\Checkmark} & \multicolumn{1}{c}{60.05 ± 9.47} & \multicolumn{1}{c}{72.87 ± 1.72} & \multicolumn{1}{c}{75.25 ± 1.19} & \multicolumn{1}{c}{61.32 ± 4.30} & \multicolumn{1}{c}{64.59 ± 2.61} & \multicolumn{1}{c}{60.77 ± 6.66} & \multicolumn{1}{c}{61.94 ± 5.11} & \multicolumn{1}{c}{66.69 ± 5.80} \\
          & \multicolumn{1}{c}{\Checkmark} & \multicolumn{1}{c|}{\XSolidBrush} & \multicolumn{1}{c}{61.53 ± 4.25} & \multicolumn{1}{c}{68.95 ± 3.27} & \multicolumn{1}{c}{73.49 ± 2.87} & \multicolumn{1}{c}{57.77 ± 4.29} & \multicolumn{1}{c}{61.32 ± 4.12} & \multicolumn{1}{c}{59.15 ± 4.22} & \multicolumn{1}{c}{59.58 ± 4.10} & \multicolumn{1}{c}{66.78 ± 3.36} \\
          & \multicolumn{1}{c}{\Checkmark} & \multicolumn{1}{c|}{\Checkmark} & \multicolumn{1}{c}{\cellcolor{pos}\textbf{68.57} ± \textbf{1.03}} & \multicolumn{1}{c}{\cellcolor{pos}\textbf{74.86} ± \textbf{0.72}} & \multicolumn{1}{c}{\cellcolor{pos}\textbf{76.57} ± \textbf{1.21}} & \multicolumn{1}{c}{\cellcolor{pos}\textbf{68.38} ± \textbf{2.16}} & \multicolumn{1}{c}{\cellcolor{pos}\textbf{69.03} ± \textbf{2.14}} & \multicolumn{1}{c}{\cellcolor{pos}\textbf{68.61} ± \textbf{1.61}} & \multicolumn{1}{c}{\cellcolor{pos}\textbf{68.59} ± \textbf{1.82}} & \multicolumn{1}{c}{\cellcolor{pos}\textbf{72.22} ± \textbf{0.79}} \\
    \midrule
    \multicolumn{1}{c}{\multirow{4}[1]{*}{System A}} & \multicolumn{1}{c}{\XSolidBrush} & \multicolumn{1}{c|}{\XSolidBrush} & \multicolumn{1}{c}{70.19 ± 4.42} & \multicolumn{1}{c}{81.05 ± 4.18} & \multicolumn{1}{c}{85.65 ± 3.36} & \multicolumn{1}{c}{70.10 ± 4.10} & \multicolumn{1}{c}{72.13 ± 4.25} & \multicolumn{1}{c}{70.01 ± 4.29} & \multicolumn{1}{c}{70.62 ± 4.26} & \multicolumn{1}{c}{77.10 ± 3.87} \\
          & \multicolumn{1}{c}{\XSolidBrush} & \multicolumn{1}{c|}{\Checkmark} & \multicolumn{1}{c}{74.78 ± 1.77} & \multicolumn{1}{c}{85.34 ± 1.60} & \multicolumn{1}{c}{88.18 ± 1.43} & \multicolumn{1}{c}{75.18 ± 1.70} & \multicolumn{1}{c}{76.79 ± 1.74} & \multicolumn{1}{c}{75.07 ± 1.64} & \multicolumn{1}{c}{75.55 ± 1.59} & \multicolumn{1}{c}{80.90 ± 0.98} \\
          & \multicolumn{1}{c}{\Checkmark} & \multicolumn{1}{c|}{\XSolidBrush} & \multicolumn{1}{c}{73.10 ± 4.55} & \multicolumn{1}{c}{84.04 ± 1.83} & \multicolumn{1}{c}{86.21 ± 1.72} & \multicolumn{1}{c}{74.53 ± 2.87} & \multicolumn{1}{c}{75.60 ± 1.72} & \multicolumn{1}{c}{73.96 ± 3.62} & \multicolumn{1}{c}{74.43 ± 2.92} & \multicolumn{1}{c}{79.23 ± 2.94} \\
          & \multicolumn{1}{c}{\Checkmark} & \multicolumn{1}{c|}{\Checkmark} & \multicolumn{1}{c}{\cellcolor{pos}\textbf{75.28} ± \textbf{2.23}} & \multicolumn{1}{c}{\cellcolor{pos}\textbf{85.60} ± \textbf{1.43}} & \multicolumn{1}{c}{\cellcolor{pos}\textbf{88.55} ± \textbf{1.15}} & \multicolumn{1}{c}{\cellcolor{pos}\textbf{76.41} ± \textbf{1.22}} & \multicolumn{1}{c}{\cellcolor{pos}\textbf{77.07} ± \textbf{1.30}} & \multicolumn{1}{c}{\cellcolor{pos}\textbf{76.13} ± \textbf{1.39}} & \multicolumn{1}{c}{\cellcolor{pos}\textbf{76.39} ± \textbf{1.21}} & \multicolumn{1}{c}{\cellcolor{pos}\textbf{81.23} ± \textbf{2.79}} \\
    \hline
    \end{tabular}%
    }
  \label{rq3:rca}%
  \vspace{-0.4cm}
\end{table*}%

\begin{table}[htbp]
  \centering
  \caption{Ablation results on the anomaly detection task.}
  \scalebox{0.85}{
    \begin{tabular}{rrrrrr}
    \hline
    \multicolumn{1}{c}{\textbf{Dataset}} & \multicolumn{1}{c}{\textbf{ILRL}} & \multicolumn{1}{c|}{\textbf{CDA}} & \multicolumn{1}{c}{\textbf{Precision}} & \multicolumn{1}{c}{\textbf{Recall}} & \multicolumn{1}{c}{\textbf{F1-Score}} \\
    \midrule
    \multicolumn{1}{c}{\multirow{4}[2]{*}{BGL}} & \multicolumn{1}{c}{\XSolidBrush} & \multicolumn{1}{c|}{\XSolidBrush} & \multicolumn{1}{c}{88.14 ± 4.59} & \multicolumn{1}{c}{98.18 ± 0.47} & \multicolumn{1}{c}{92.82 ± 2.39} \\
          & \multicolumn{1}{c}{\XSolidBrush} & \multicolumn{1}{c|}{\Checkmark} & \multicolumn{1}{c}{90.04 ± 1.98} & \multicolumn{1}{c}{97.99 ± 0.27} & \multicolumn{1}{c}{93.84 ± 1.15} \\
          & \multicolumn{1}{c}{\Checkmark} & \multicolumn{1}{c|}{\XSolidBrush} & \multicolumn{1}{c}{89.37 ± 1.69} & \multicolumn{1}{c}{98.09 ± 0.34} & \multicolumn{1}{c}{93.52 ± 0.80} \\
          & \multicolumn{1}{c}{\Checkmark} & \multicolumn{1}{c|}{\Checkmark} & \multicolumn{1}{c}{91.22 ± 1.90} & \multicolumn{1}{c}{97.37 ± 0.36} & \multicolumn{1}{c}{\cellcolor{pos}\textbf{94.19} ± \textbf{1.05}} \\
    \midrule
    \multicolumn{1}{c}{\multirow{4}[2]{*}{Thunderbird}} & \multicolumn{1}{c}{\XSolidBrush} & \multicolumn{1}{c|}{\XSolidBrush} & \multicolumn{1}{c}{97.82 ± 0.53} & \multicolumn{1}{c}{78.74 ± 1.09} & \multicolumn{1}{c}{87.25 ± 0.62} \\
          & \multicolumn{1}{c}{\XSolidBrush} & \multicolumn{1}{c|}{\Checkmark} & \multicolumn{1}{c}{98.36 ± 0.85} & \multicolumn{1}{c}{79.15 ± 0.50} & \multicolumn{1}{c}{87.71 ± 0.25} \\
          & \multicolumn{1}{c}{\Checkmark} & \multicolumn{1}{c|}{\XSolidBrush} & \multicolumn{1}{c}{98.90 ± 0.61} & \multicolumn{1}{c}{79.26 ± 0.44} & \multicolumn{1}{c}{88.00 ± 0.45} \\
          & \multicolumn{1}{c}{\Checkmark} & \multicolumn{1}{c|}{\Checkmark} & \multicolumn{1}{c}{98.08 ± 0.75} & \multicolumn{1}{c}{82.82 ± 2.18} & \multicolumn{1}{c}{\cellcolor{pos}\textbf{89.80} ± \textbf{1.59}} \\
    \midrule
    \multicolumn{1}{c}{\multirow{4}[1]{*}{System A}} & \multicolumn{1}{c}{\XSolidBrush} & \multicolumn{1}{c|}{\XSolidBrush} & \multicolumn{1}{c}{77.19 ± 2.38} & \multicolumn{1}{c}{94.34 ± 1.22} & \multicolumn{1}{c}{84.87 ± 1.00} \\
          & \multicolumn{1}{c}{\XSolidBrush} & \multicolumn{1}{c|}{\Checkmark} & \multicolumn{1}{c}{79.06 ± 2.38} & \multicolumn{1}{c}{94.23 ± 1.99} & \multicolumn{1}{c}{85.94 ± 1.26} \\
          & \multicolumn{1}{c}{\Checkmark} & \multicolumn{1}{c|}{\XSolidBrush} & \multicolumn{1}{c}{81.69 ± 5.84} & \multicolumn{1}{c}{91.85 ± 3.97} & \multicolumn{1}{c}{86.19 ± 1.45} \\
          & \multicolumn{1}{c}{\Checkmark} & \multicolumn{1}{c|}{\Checkmark} & \multicolumn{1}{c}{82.75 ± 1.02} & \multicolumn{1}{c}{93.30 ± 1.57} & \multicolumn{1}{c}{\cellcolor{pos}\textbf{87.70} ± \textbf{0.58}} \\
    \hline
    \end{tabular}%
    }
  \label{rq3:ad}%
  \vspace{-0.4cm}
\end{table}%

\subsection{RQ2: How effective is Chimera in the log-based fault diagnosis?}
This RQ evaluates the effectiveness of Chimera in log-based fault diagnosis tasks and highlights the importance of deploying end-to-end fault diagnosis systems using a task-interactive strategy. To this end, we compared Chimera with state-of-the-art log-based fault diagnosis methods that are deployed using a task-independent strategy, including those constructed with pipeline and end-to-end paradigms. Among these, RobustFlagger is a fault diagnosis method deployed in a pipeline manner, consisting of the state-of-the-art anomaly detection method LogRobust \cite{zhang2019robust} and the leading root cause localization method LogFaultFlagger \cite{amar2019mining} from RQ1. Tables \ref{rq2:rca} and \ref{rq2:ad} present our results. 

Generally speaking, Chimera achieved the best performance across three datasets. Compared to the advanced task-independent method built on a pipeline paradigm, Chimera's results for the anomaly detection exceeded the F1 scores by 3.99\%, 4.31\%, and 2.92\%, respectively. For the root cause localization, Chimera's results averaged 29.30\%, 33.61\%, 37.09\%, and 30.58\% higher in HR@k, PR@k, MAP@k, and MRR, respectively. 
Task-independent methods constructed using a pipeline paradigm perform detection and localization completely independently, resulting in the accumulation of diagnostic bias within the system. 
In contrast, Chimera implements end-to-end fault diagnosis through interactive learning between the two tasks, effectively addressing diagnostic bias. 
Compared to the advanced task-independent methods built on an end-to-end paradigm, Chimera's results for the anomaly detection increased the F1 scores by 4.24\%, 5.00\%, and 4.66\%, respectively. For the root cause localization, Chimera's results averaged 19.24\%, 14.75\%, 21.05\%, and 21.20\% higher in HR@k, PR@k, MAP@k, and MRR, respectively. 
Existing end-to-end method deploy by simply sharing features between the two tasks, neglecting the utilization of diagnostic information.
In contrast, Chimera utilizes interactive learning between the two tasks to facilitate knowledge transfer of diagnostic information, achieving more effective fault diagnosis. 
% This evidence demonstrates the effectiveness of Chimera in log-based fault diagnosis tasks and underscore the importance of deploying end-to-end fault diagnosis systems using a task-interactive strategy.

\subsection{RQ3: How do different modules contribute to Chimera?}
In this section, we perform an ablation study to assess the effectiveness of two critical components in Chimera: Interactive Log Representation Learning (ILRL) and Cross-granularity Diagnostic Alignment (CDA). 
% The former engages in interaction at the feature level, whereas the latter interacts at the diagnostic results level. 
The findings of the ablation study are shown in Tables \ref{rq3:rca} and \ref{rq3:ad}. 

In general, the ablation of all components resulted in different levels of performance degradation in both the anomaly detection and root cause localization tasks, confirming the effectiveness of the components we proposed. During the ablation of the ILRL component, Chimera's results for the anomaly detection task were 2.09\% higher in terms of F1 scores. In the root cause localization task, Chimera achieved results that were 8.52\%, 7.06\%, 7.84\%, and 5.53\% higher in HR@k, PR@k, MAP@k, and MRR, respectively. This suggests that moderate interaction between the two tasks at the feature level is advantageous, aiding in facilitating feature knowledge sharing between the tasks. 

During the ablation of the CDA component, Chimera's results were 1.51\% higher in F1 scores for the anomaly detection task. In the root cause localization task, Chimera achieved results that were 7.04\%, 10.61\%, 9.46\%, and 5.44\% higher in HR@k, PR@k, MAP@k, and MRR, respectively. This suggests that interaction at the diagnostic results level effectively capitalizes on the insights of the two sub-tasks across different diagnostic granularities, facilitating more efficient fault diagnosis. The concurrent ablation of both components yielded notably greater benefits, with F1 scores in the anomaly detection task exceeding by 2.83\% and MRR in the root cause localization task exceeding by 8.31\%. 
These findings indicate that both the ILRL and CDA facilitate bidirectional interaction and knowledge transfer between the two sub-tasks, enabling effective fault diagnosis.

\subsection{RQ4: How effective is Chimera in addressing diagnostic bias?}
This RQ evaluates whether Chimera effectively addresses diagnostic bias in fault diagnosis. To this end, we followed the setup in Section \ref{sec:bias} and further compared the root cause localization performance of various fault diagnosis methods under theoretical and actual anomaly detection settings. The performance was calculated using the average HR@1, HR@3, and HR@5 metrics, with results shown in table \ref{exp_bias}.

% Table generated by Excel2LaTeX from sheet 'Add_res'
\begin{table}[htbp]
  \centering
  \caption{Comparison results with baseline methods for addressing diagnostic bias.}
    \scalebox{0.85}{
    \begin{tabular}{rrrrr}
    \hline
    \multicolumn{1}{c}{\textbf{Dataset}} & \multicolumn{1}{c}{\textbf{Model}} & \multicolumn{1}{c}{\textbf{Theoretical}} & \multicolumn{1}{c}{\textbf{Actual}} & \multicolumn{1}{c}{\textbf{Bias}} \\
    \midrule
    \multicolumn{1}{c}{\multirow{5}[4]{*}{BGL}} & \multicolumn{1}{c}{RobustFlagger} & \multicolumn{1}{c}{60.42} & \multicolumn{1}{c}{59.21} & \multicolumn{1}{c}{-1.21} \\
          & \multicolumn{1}{c}{LogRCA} & \multicolumn{1}{c}{63.07} & \multicolumn{1}{c}{59.69} & \multicolumn{1}{c}{-3.38} \\
          & \multicolumn{1}{c}{AFAFormer} & \multicolumn{1}{c}{84.96} & \multicolumn{1}{c}{81.35} & \multicolumn{1}{c}{-3.61} \\
          & \multicolumn{1}{c}{Eadro} & \multicolumn{1}{c}{64.67} & \multicolumn{1}{c}{56.80} & \multicolumn{1}{c}{-7.87} \\
         & \multicolumn{1}{c}{Chimera} & \multicolumn{1}{c}{\cellcolor{pos}\textbf{94.39}} & \multicolumn{1}{c}{\cellcolor{pos}\textbf{93.26}} & \multicolumn{1}{c}{\cellcolor{pos}\textbf{-1.13}} \\
    \midrule
    \multicolumn{1}{c}{\multirow{5}[4]{*}{Thunderbird}} & \multicolumn{1}{c}{RobustFlagger} & \multicolumn{1}{c}{37.96} & \multicolumn{1}{c}{22.26} & \multicolumn{1}{c}{-15.70} \\
          & \multicolumn{1}{c}{LogRCA} & \multicolumn{1}{c}{42.24} & \multicolumn{1}{c}{35.89} & \multicolumn{1}{c}{-6.35} \\
          & \multicolumn{1}{c}{AFAFormer} & \multicolumn{1}{c}{42.89} & \multicolumn{1}{c}{35.62} & \multicolumn{1}{c}{-7.27} \\
          & \multicolumn{1}{c}{Eadro} & \multicolumn{1}{c}{46.60} & \multicolumn{1}{c}{29.28} & \multicolumn{1}{c}{-17.32} \\
         & \multicolumn{1}{c}{Chimera} & \multicolumn{1}{c}{\cellcolor{pos}\textbf{76.74}} & \multicolumn{1}{c}{\cellcolor{pos}\textbf{73.33}} & \multicolumn{1}{c}{\cellcolor{pos}\textbf{-3.41}} \\
    \midrule
    \multicolumn{1}{c}{\multirow{5}[4]{*}{System A}} & \multicolumn{1}{c}{RobustFlagger} & \multicolumn{1}{c}{66.41} & \multicolumn{1}{c}{63.13} & \multicolumn{1}{c}{-3.28} \\
          & \multicolumn{1}{c}{LogRCA} & \multicolumn{1}{c}{70.39} & \multicolumn{1}{c}{66.06} & \multicolumn{1}{c}{-4.33} \\
          & \multicolumn{1}{c}{AFAFormer} & \multicolumn{1}{c}{78.51} & \multicolumn{1}{c}{75.04} & \multicolumn{1}{c}{-3.47} \\
          & \multicolumn{1}{c}{Eadro} & \multicolumn{1}{c}{71.67} & \multicolumn{1}{c}{61.38} & \multicolumn{1}{c}{-10.29} \\
         & \multicolumn{1}{c}{Chimera} & \multicolumn{1}{c}{\cellcolor{pos}\textbf{85.38}} & \multicolumn{1}{c}{\cellcolor{pos}\textbf{83.14}} & \multicolumn{1}{c}{\cellcolor{pos}\textbf{-2.24}} \\
    \hline
    \end{tabular}%
    }
  \label{exp_bias}%
  \vspace{-0.4cm}
\end{table}%

Overall, Chimera effectively reduced the adverse effects of diagnostic bias, achieving the best performance. Existing methods conduct fault diagnosis in a task-independent manner, performing anomaly detection and root cause localization independently, which results in inaccurate anomaly detection severely affecting the root cause localization task, reducing performance by up to 17.32\%. Even the best methods caused a diagnostic bias of 6.35\% when handling the Thunderbird dataset, which is catastrophic for real-world fault diagnosis. In contrast, Chimera achieved fault diagnosis through bidirectional interaction and knowledge transfer between the anomaly detection and root cause localization tasks, resulting in a maximum diagnostic bias of only 3.41\% within a unified end-to-end framework. This demonstrates that Chimera is indeed capable of effectively handling diagnostic bias.

\subsection{RQ5: How effective is Chimera in addressing suboptimal diagnosis?}
\begin{figure}[htbp]
\centerline{
\includegraphics[width=0.9\linewidth]{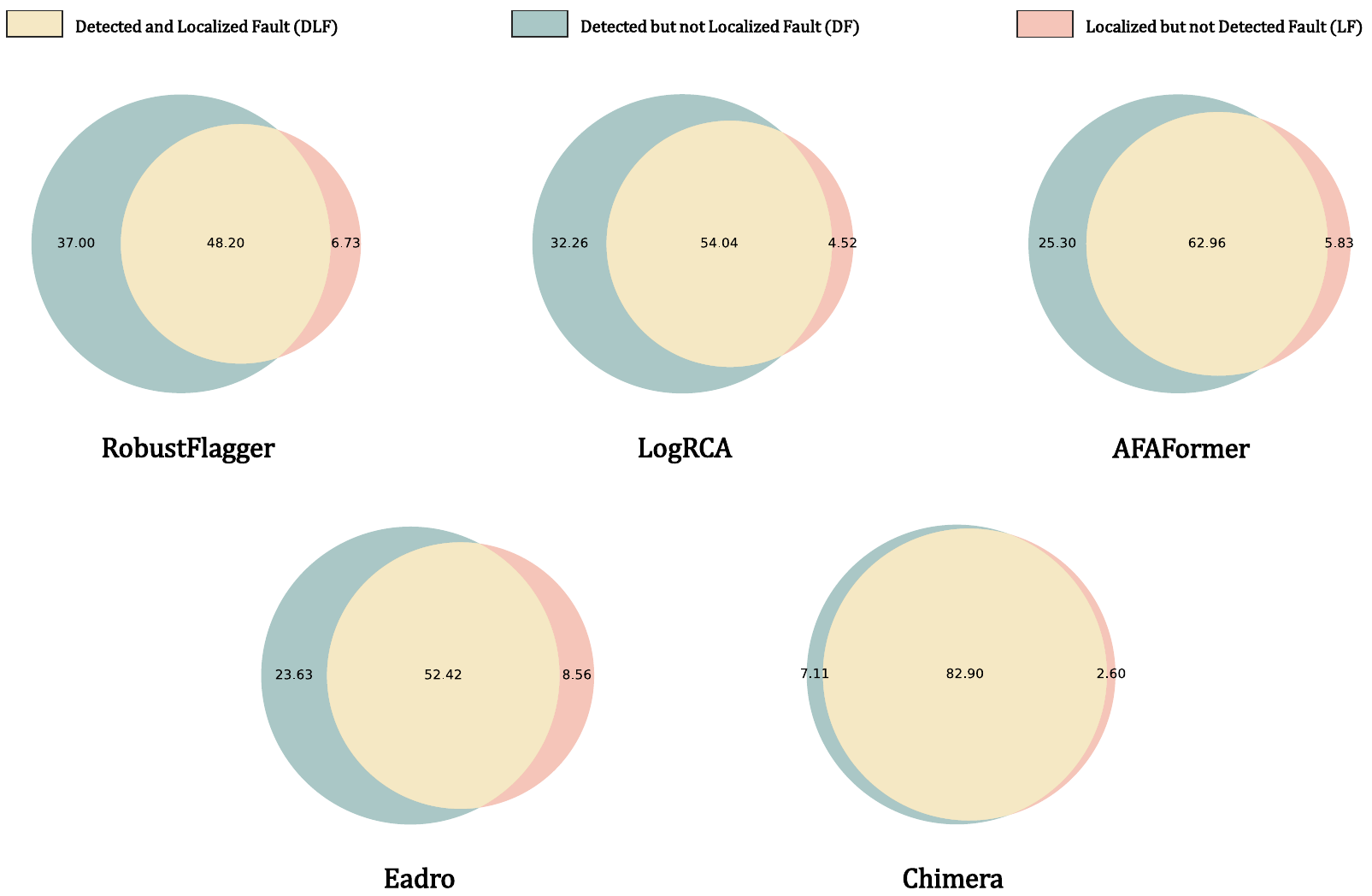}
}
\caption{Comparison results with baselines for addressing suboptimal diagnosis. 
% The green portion represents Detected but not Localized Fault (DF), the red portion represents Localized but not Detected Fault (LF), and the yellow portion represents Detected and Localized Fault (DLF).
}
\label{coll_all}
\vspace{-0.4cm}
\end{figure}

This RQ evaluates whether Chimera effectively handles suboptimal diagnoses. To this end, we followed the setup in Section \ref{sec:empirical_diag} and further compared the proportions of suboptimal diagnoses produced by various fault diagnosis methods. The evaluation results shown in figure \ref{coll_all}.
% Specifically, we examined the proportions of DF, LF, and DLF produced by existing methods and the proposed method when handling BGL, Thunderbird, and System A, and reported the averages across the three datasets, with the evaluation results shown in figure \ref{coll_all}. 

Overall, Chimera produced the fewest suboptimal diagnoses, with the majority of faults being detected and localized simultaneously. Existing methods overlook the collaborative relationship between the two tasks, which results in system faults not being detected and localized simultaneously, leading to the occurrence of suboptimal diagnoses, averaging 43.73\%, 36.78\%, 31.13\%, and 32.19\%. In contrast, Chimera produced only 9.71\% suboptimal diagnoses, reducing this by 21.42\% compared to the advanced methods. Furthermore, Chimera generated the highest number of optimal diagnoses (DLF), exceeding the advanced methods by 19.94\%. This success is attributed to Chimera's thoughtfully designed interactive learning strategy, which facilitates knowledge transfer between anomaly detection and root cause localization, leveraging different insights from diagnostic tasks for fault diagnosis. 
% This evidence strongly indicates that Chimera can effectively transform suboptimal diagnoses into optimal ones, thereby achieving more effective fault diagnosis.

% \subsection{Run Time Analysis}
% We analyzed the runtime of Chimera in fault diagnosis using a single NVIDIA RTX 3090 to demonstrate that Chimera can be easily applied to real industrial fault diagnosis, with the results of the runtime analysis shown in table \ref{runtime}. 

% \input{tabs/runtime}

% Among the baseline methods, RobustFlagger achieved the fastest diagnostic speed, attributed to its use of simple semantic clustering for root cause localization, but at the cost of suboptimal diagnostic performance. In contrast, Chimera achieved a suboptimal diagnostic speed, significantly surpassing other baseline methods. Existing methods are either implemented based on complex neural network architectures or predict root cause scores by directly processing log sequences, which increases the overhead of the diagnostic process. In contrast, Chimera utilizes a lightweight structure as its network backbone and employs a multi-instance learning paradigm to predict root cause scores through batch processing of log entries, resulting in faster diagnostic speeds. This evidence demonstrates that, compared to existing methods, Chimera exhibits efficient runtime performance, making it suitable for real industrial fault diagnosis.

\section{Related Work}
\subsection{Log-based Fault Diagnosis}
Analyzing logs to discover and identify issues has long been an active area of research \cite{jia2024hilogx,duan2023aclog,du2017deeplog,meng2019loganomaly,zhang2019robust,icsme2020,icse2020,logflash,kddcfg,logsed}. Recent work in fault diagnosis has focused on two tasks: anomaly detection and root cause localization. Some state-of-the-art anomaly detection methods \cite{jia2024hilogx,logflash,kddcfg,logsed,icwslogsed} first extract event sequences and then generate graph-based models to compare with log sequences in production environments to detect conflicts. Other methods \cite{yang2021semi,duan2023aclog,meng2019loganomaly,zhang2019robust,icsme2020,icse2020} typically construct deep learning-based models to capture sequential features of log events. Some other works \cite{wittkopp2024logrca, hou2021diagnosing, huo2023evlog, zhang2021onion, he2018identifying, amar2019mining, lee2023eadro} focus on locating the root causes of anomalies. LogRCA \cite{wittkopp2024logrca} is a root cause localization method that ranks all logs within an investigation time window using the principles of PU Learning. Onion \cite{zhang2021onion} proposed a log aggregation form called Log Clique to identify event-indicative logs. Building on these works, researchers aim to propose a unified fault diagnosis framework. Eadro \cite{lee2023eadro} is an end-to-end fault diagnosis method designed for microservices systems that aggregates multi-source monitoring data by modeling internal service behaviors. However, these works overlook the interaction between anomaly detection and root cause localization, failing to bridge the gap in data forms and diagnostic objectives, which leads to issues such as the propagation of diagnostic bias. 
% In contrast to these works, Chimera is based on interactive multi-task learning, meticulously designing interaction strategies for anomaly detection and root cause localization at the levels of data, features, and diagnostic results, to interactively achieve these two sub-tasks within a unified end-to-end framework.

\subsection{Multi-task Learning}
Multi-task learning (MTL) is a machine learning paradigm designed to utilize valuable information from multiple related tasks to improve the generalization performance across all tasks \cite{zhang2021survey, standley2020tasks, vandenhende2021multi, misra2016cross, ruder2019latent, long2017learning}. Recent studies on MTL have centered on the interactions of tasks related to features. ASAP \cite{bu2021asap} was the first to investigate an end-to-end approach for extracting aspects and sentiments within a MTL framework, achieving deployment by sharing task features. SPRM \cite{lin2020shared} and HI-ASA \cite{chen2022hierarchical} explore interaction strategies focused on features, with the goal of capturing different levels of feature correlation among various sentences. DMVAE \cite{lee2021private} introduces an innovative disentangled VAE approach intended to segregate the private and shared latent spaces in multimodal contexts. Considering that feature interactions are independent of specific tasks, additional research has focused on explicit interactions among tasks. IMN \cite{he2019interactive} utilizes an iterative message-passing framework to model interactions between tasks, with the goal of enhancing the transfer of information. STAGE \cite{liang2023stage} introduces a greedy reasoning strategy for triplet extraction that accounts for the mutual span constraints between sentiment fragments and aspects. In contrast to these works, Chimera concurrently establishes implicit feature interactions and explicit task interactions, utilizing a variety of interaction strategies to promote knowledge transfer for anomaly detection and root cause analysis.
\section{Conclusion}
In this paper, we propose an end-to-end log-based fault diagnosis method, called Chimera. The key idea of Chimera is to achieve end-to-end fault diagnosis through bidirectional interaction and knowledge transfer between anomaly detection and root cause localization. It is built upon interactive multi-task learning, where interaction strategies for anomaly detection and root cause localization are meticulously designed at the levels of data, features, and diagnostic results, enabling these two sub-tasks to interact within a unified end-to-end framework. Evaluation on two public datasets and one industrial dataset shows that Chimera achieves advantages of over 2.92\% to 5.00\% in anomaly detection and over 19.01\% to 37.09\% in root cause localization compared to existing methods. 
Chimera has been successfully deployed in production, serving an industrial cloud platform.

\section*{Acknowledgment}
This work was supported by National Key Laboratory of Data Space Technology and System.

\bibliographystyle{IEEEtran}
\bibliography{sample}

\begin{thebibliography}{10}
\providecommand{\url}[1]{#1}
\csname url@samestyle\endcsname
\providecommand{\newblock}{\relax}
\providecommand{\bibinfo}[2]{#2}
\providecommand{\BIBentrySTDinterwordspacing}{\spaceskip=0pt\relax}
\providecommand{\BIBentryALTinterwordstretchfactor}{4}
\providecommand{\BIBentryALTinterwordspacing}{\spaceskip=\fontdimen2\font plus
\BIBentryALTinterwordstretchfactor\fontdimen3\font minus \fontdimen4\font\relax}
\providecommand{\BIBforeignlanguage}[2]{{%
\expandafter\ifx\csname l@#1\endcsname\relax
\typeout{** WARNING: IEEEtran.bst: No hyphenation pattern has been}%
\typeout{** loaded for the language `#1'. Using the pattern for}%
\typeout{** the default language instead.}%
\else
\language=\csname l@#1\endcsname
\fi
#2}}
\providecommand{\BIBdecl}{\relax}
\BIBdecl

\bibitem{niu2023locating}
S.~Niu, J.~Jin, X.~Huang, Y.~Wang, W.~Xu, and Y.~Kong, ``Locating faulty applications via semantic and topology estimation,'' in \emph{Companion Proceedings of the ACM Web Conference 2023}, 2023, pp. 528--532.

\bibitem{he2025execoder}
M.~He, F.~Yang, P.~Zhao, W.~Yin, Y.~Kang, Q.~Lin, S.~Rajmohan, D.~Zhang, and Q.~Zhang, ``Execoder: Empowering large language models with executability representation for code translation,'' \emph{arXiv preprint arXiv:2501.18460}, 2025.

\bibitem{lin2023edits}
Q.~Lin, T.~Li, P.~Zhao, Y.~Liu, M.~Ma, L.~Zheng, M.~Chintalapati, B.~Liu, P.~Wang, H.~Zhang \emph{et~al.}, ``Edits: An easy-to-difficult training strategy for cloud failure prediction,'' in \emph{Companion Proceedings of the ACM Web Conference 2023}, 2023, pp. 371--375.

\bibitem{chen2025warriormath}
Y.~Chen, M.~He, F.~Yang, P.~Zhao, L.~Wang, Y.~Kang, Y.~Dong, Y.~Zhan, H.~Sun, Q.~Lin \emph{et~al.}, ``Warriormath: Enhancing the mathematical ability of large language models with a defect-aware framework,'' \emph{arXiv preprint arXiv:2508.01245}, 2025.

\bibitem{li2023elastic}
Y.~Li, H.~Yuan, Z.~Fu, X.~Ma, M.~Xu, and S.~Wang, ``Elastic: edge workload forecasting based on collaborative cloud-edge deep learning,'' in \emph{Proceedings of the ACM Web Conference 2023}, 2023, pp. 3056--3066.

\bibitem{luo2021ntam}
C.~Luo, P.~Zhao, B.~Qiao, Y.~Wu, H.~Zhang, W.~Wu, W.~Lu, Y.~Dang, S.~Rajmohan, Q.~Lin \emph{et~al.}, ``Ntam: neighborhood-temporal attention model for disk failure prediction in cloud platforms,'' in \emph{Proceedings of the Web Conference 2021}, 2021, pp. 1181--1191.

\bibitem{llmelog}
M.~He, T.~Jia, C.~Duan, H.~Cai, Y.~Li, and G.~Huang, ``Llmelog: An approach for anomaly detection based on llm-enriched log events,'' in \emph{2024 IEEE 35th International Symposium on Software Reliability Engineering (ISSRE)}.\hskip 1em plus 0.5em minus 0.4em\relax IEEE, 2024, pp. 132--143.

\bibitem{huang2024demystifying}
J.~Huang, Z.~Jiang, J.~Liu, Y.~Huo, J.~Gu, Z.~Chen, C.~Feng, H.~Dong, Z.~Yang, and M.~R. Lyu, ``Demystifying and extracting fault-indicating information from logs for failure diagnosis,'' in \emph{2024 IEEE 35th International Symposium on Software Reliability Engineering (ISSRE)}.\hskip 1em plus 0.5em minus 0.4em\relax IEEE, 2024, pp. 511--522.

\bibitem{chen2024lara}
F.~Chen, Z.~Qin, M.~Zhou, Y.~Zhang, S.~Deng, L.~Fan, G.~Pang, and Q.~Wen, ``Lara: A light and anti-overfitting retraining approach for unsupervised time series anomaly detection,'' in \emph{Proceedings of the ACM on Web Conference 2024}, 2024, pp. 4138--4149.

\bibitem{tan2024air}
Y.~Tan, Q.~Li, J.~Peng, Z.~Yuan, and Y.~Jiang, ``Air-cad: Edge-assisted multi-drone network for real-time crowd anomaly detection,'' in \emph{Proceedings of the ACM on Web Conference 2024}, 2024, pp. 2817--2825.

\bibitem{yu2024supervised}
Z.~Yu, S.~Zhang, M.~Sun, Y.~Li, Y.~Zhao, X.~Hua, L.~Zhu, X.~Wen, and D.~Pei, ``Supervised fine-tuning for unsupervised kpi anomaly detection for mobile web systems,'' in \emph{Proceedings of the ACM on Web Conference 2024}, 2024, pp. 2859--2869.

\bibitem{dai2021sdfvae}
L.~Dai, T.~Lin, C.~Liu, B.~Jiang, Y.~Liu, Z.~Xu, and Z.-L. Zhang, ``Sdfvae: Static and dynamic factorized vae for anomaly detection of multivariate cdn kpis,'' in \emph{Proceedings of the Web Conference 2021}, 2021, pp. 3076--3086.

\bibitem{huang2022semi}
T.~Huang, P.~Chen, and R.~Li, ``A semi-supervised vae based active anomaly detection framework in multivariate time series for online systems,'' in \emph{Proceedings of the ACM Web Conference 2022}, 2022, pp. 1797--1806.

\bibitem{wong2016survey}
W.~E. Wong, R.~Gao, Y.~Li, R.~Abreu, and F.~Wotawa, ``A survey on software fault localization,'' \emph{IEEE Transactions on Software Engineering}, vol.~42, no.~8, pp. 707--740, 2016.

\bibitem{zhang2024failure}
S.~Zhang, S.~Xia, W.~Fan, B.~Shi, X.~Xiong, Z.~Zhong, M.~Ma, Y.~Sun, and D.~Pei, ``Failure diagnosis in microservice systems: A comprehensive survey and analysis,'' \emph{arXiv preprint arXiv:2407.01710}, 2024.

\bibitem{jiang2011efficient}
M.~Jiang, M.~A. Munawar, T.~Reidemeister, and P.~A. Ward, ``Efficient fault detection and diagnosis in complex software systems with information-theoretic monitoring,'' \emph{IEEE Transactions on Dependable and Secure Computing}, vol.~8, no.~4, pp. 510--522, 2011.

\bibitem{zhang2019robust}
X.~Zhang, Y.~Xu, Q.~Lin, B.~Qiao, H.~Zhang, Y.~Dang, C.~Xie, X.~Yang, Q.~Cheng, Z.~Li \emph{et~al.}, ``Robust log-based anomaly detection on unstable log data,'' in \emph{Proceedings of the 2019 27th ACM joint meeting on European software engineering conference and symposium on the foundations of software engineering}, 2019, pp. 807--817.

\bibitem{xia2019bugidentifier}
W.~Xia, Y.~Li, T.~Jia, and Z.~Wu, ``Bugidentifier: An approach to identifying bugs via log mining for accelerating bug reporting stage,'' in \emph{2019 IEEE 19th International Conference on Software Quality, Reliability and Security (QRS)}.\hskip 1em plus 0.5em minus 0.4em\relax IEEE, 2019, pp. 167--175.

\bibitem{sun2025exploring}
J.~Sun, T.~Jia, M.~He, Y.~Wu, Y.~Li, and G.~Huang, ``Exploring variable potential for llm-based log parsing efficiency and reduced costs,'' in \emph{Proceedings of the 33rd ACM International Conference on the Foundations of Software Engineering}, 2025, pp. 596--600.

\bibitem{zhao2025few}
X.~Zhao, T.~Jia, M.~He, Y.~Wu, Y.~Li, and G.~Huang, ``From few-label to zero-label: An approach for cross-system log-based anomaly detection with meta-learning,'' in \emph{Proceedings of the 33rd ACM International Conference on the Foundations of Software Engineering}, 2025, pp. 661--665.

\bibitem{zhang2025survey}
L.~Zhang, L.~Fang, C.~Duan, M.~He, L.~Pan, P.~Xiao, S.~Huang, Y.~Zhai, X.~Hu, P.~S. Yu \emph{et~al.}, ``A survey on parallel text generation: From parallel decoding to diffusion language models,'' \emph{arXiv preprint arXiv:2508.08712}, 2025.

\bibitem{reidemeister2010identifying}
T.~Reidemeister, M.~A. Munawar, and P.~A. Ward, ``Identifying symptoms of recurrent faults in log files of distributed information systems,'' in \emph{2010 IEEE Network Operations and Management Symposium-NOMS 2010}.\hskip 1em plus 0.5em minus 0.4em\relax IEEE, 2010, pp. 187--194.

\bibitem{tong2016approach}
J.~Tong, L.~Ying, T.~Hongyan, and W.~Zhonghai, ``An approach to pinpointing bug-induced failure in logs of open cloud platforms,'' in \emph{2016 IEEE 9th International Conference on Cloud Computing (CLOUD)}.\hskip 1em plus 0.5em minus 0.4em\relax IEEE, 2016, pp. 294--302.

\bibitem{yang2021semi}
L.~Yang, J.~Chen, Z.~Wang, W.~Wang, J.~Jiang, X.~Dong, and W.~Zhang, ``Semi-supervised log-based anomaly detection via probabilistic label estimation,'' in \emph{2021 IEEE/ACM 43rd International Conference on Software Engineering (ICSE)}.\hskip 1em plus 0.5em minus 0.4em\relax IEEE, 2021, pp. 1448--1460.

\bibitem{du2017deeplog}
M.~Du, F.~Li, G.~Zheng, and V.~Srikumar, ``Deeplog: Anomaly detection and diagnosis from system logs through deep learning,'' in \emph{Proceedings of the 2017 ACM SIGSAC conference on computer and communications security}, 2017, pp. 1285--1298.

\bibitem{meng2019loganomaly}
W.~Meng, Y.~Liu, Y.~Zhu, S.~Zhang, D.~Pei, Y.~Liu, Y.~Chen, R.~Zhang, S.~Tao, P.~Sun \emph{et~al.}, ``Loganomaly: Unsupervised detection of sequential and quantitative anomalies in unstructured logs.'' in \emph{IJCAI}, vol.~19, no.~7, 2019, pp. 4739--4745.

\bibitem{jiang2025l4}
Z.~Jiang, J.~Huang, G.~Yu, Z.~Chen, Y.~Li, R.~Zhong, C.~Feng, Y.~Yang, Z.~Yang, and M.~Lyu, ``L4: Diagnosing large-scale llm training failures via automated log analysis,'' in \emph{Proceedings of the 33rd ACM International Conference on the Foundations of Software Engineering}, 2025, pp. 51--63.

\bibitem{huang2025no}
J.~Huang, Z.~Jiang, Z.~Chen, and M.~Lyu, ``No more labelled examples? an unsupervised log parser with llms,'' \emph{Proceedings of the ACM on Software Engineering}, vol.~2, no. FSE, pp. 2406--2429, 2025.

\bibitem{zhang2024metalog}
C.~Zhang, T.~Jia, G.~Shen, P.~Zhu, and Y.~Li, ``Metalog: Generalizable cross-system anomaly detection from logs with meta-learning,'' in \emph{Proceedings of the IEEE/ACM 46th International Conference on Software Engineering}, 2024, pp. 1--12.

\bibitem{jia2024hilogx}
T.~Jia, Y.~Li, Y.~Yang, and G.~Huang, ``Hilogx: noise-aware log-based anomaly detection with human feedback,'' \emph{The VLDB Journal}, vol.~33, no.~3, pp. 883--900, 2024.

\bibitem{zhang2021cloudrca}
Y.~Zhang, Z.~Guan, H.~Qian, L.~Xu, H.~Liu, Q.~Wen, L.~Sun, J.~Jiang, L.~Fan, and M.~Ke, ``Cloudrca: A root cause analysis framework for cloud computing platforms,'' in \emph{Proceedings of the 30th ACM International Conference on Information \& Knowledge Management}, 2021, pp. 4373--4382.

\bibitem{wittkopp2024logrca}
T.~Wittkopp, P.~Wiesner, and O.~Kao, ``Logrca: Log-based root cause analysis for distributed services,'' in \emph{European Conference on Parallel Processing}.\hskip 1em plus 0.5em minus 0.4em\relax Springer, 2024, pp. 362--376.

\bibitem{hou2021diagnosing}
C.~Hou, T.~Jia, Y.~Wu, Y.~Li, and J.~Han, ``Diagnosing performance issues in microservices with heterogeneous data source,'' in \emph{2021 IEEE Intl Conf on Parallel \& Distributed Processing with Applications, Big Data \& Cloud Computing, Sustainable Computing \& Communications, Social Computing \& Networking (ISPA/BDCloud/SocialCom/SustainCom)}.\hskip 1em plus 0.5em minus 0.4em\relax IEEE, 2021, pp. 493--500.

\bibitem{huo2023evlog}
Y.~Huo, C.~Lee, Y.~Su, S.~Shan, J.~Liu, and M.~R. Lyu, ``Evlog: Identifying anomalous logs over software evolution,'' in \emph{2023 IEEE 34th International Symposium on Software Reliability Engineering (ISSRE)}.\hskip 1em plus 0.5em minus 0.4em\relax IEEE, 2023, pp. 391--402.

\bibitem{zhang2021onion}
X.~Zhang, Y.~Xu, S.~Qin, S.~He, B.~Qiao, Z.~Li, H.~Zhang, X.~Li, Y.~Dang, Q.~Lin \emph{et~al.}, ``Onion: identifying incident-indicating logs for cloud systems,'' in \emph{Proceedings of the 29th ACM Joint Meeting on European Software Engineering Conference and Symposium on the Foundations of Software Engineering}, 2021, pp. 1253--1263.

\bibitem{he2018identifying}
S.~He, Q.~Lin, J.-G. Lou, H.~Zhang, M.~R. Lyu, and D.~Zhang, ``Identifying impactful service system problems via log analysis,'' in \emph{Proceedings of the 2018 26th ACM joint meeting on European software engineering conference and symposium on the foundations of software engineering}, 2018, pp. 60--70.

\bibitem{amar2019mining}
A.~Amar and P.~C. Rigby, ``Mining historical test logs to predict bugs and localize faults in the test logs,'' in \emph{2019 IEEE/ACM 41st International Conference on Software Engineering (ICSE)}.\hskip 1em plus 0.5em minus 0.4em\relax IEEE, 2019, pp. 140--151.

\bibitem{lee2023eadro}
C.~Lee, T.~Yang, Z.~Chen, Y.~Su, and M.~R. Lyu, ``Eadro: An end-to-end troubleshooting framework for microservices on multi-source data,'' in \emph{2023 IEEE/ACM 45th International Conference on Software Engineering (ICSE)}.\hskip 1em plus 0.5em minus 0.4em\relax IEEE, 2023, pp. 1750--1762.

\bibitem{notaro2023logrule}
P.~Notaro, S.~Haeri, J.~Cardoso, and M.~Gerndt, ``Logrule: Efficient structured log mining for root cause analysis,'' \emph{IEEE Transactions on Network and Service Management}, vol.~20, no.~4, pp. 4231--4243, 2023.

\bibitem{wang2020root}
L.~Wang, N.~Zhao, J.~Chen, P.~Li, W.~Zhang, and K.~Sui, ``Root-cause metric location for microservice systems via log anomaly detection,'' in \emph{2020 IEEE international conference on web services (ICWS)}.\hskip 1em plus 0.5em minus 0.4em\relax IEEE, 2020, pp. 142--150.

\bibitem{yang2018nanolog}
S.~Yang, S.~J. Park, and J.~Ousterhout, ``$\{$NanoLog$\}$: A nanosecond scale logging system,'' in \emph{2018 USENIX Annual Technical Conference (USENIX ATC 18)}, 2018, pp. 335--350.

\bibitem{luo2018troubleshooting}
L.~Luo, S.~Nath, L.~R. Sivalingam, M.~Musuvathi, and L.~Ceze, ``Troubleshooting $\{$Transiently-Recurring$\}$ errors in production systems with $\{$Blame-Proportional$\}$ logging,'' in \emph{2018 USENIX Annual Technical Conference (USENIX ATC 18)}, 2018, pp. 321--334.

\bibitem{duan2023aclog}
C.~Duan, T.~Jia, Y.~Li, and G.~Huang, ``Aclog: An approach to detecting anomalies from system logs with active learning,'' in \emph{2023 IEEE International Conference on Web Services (ICWS)}.\hskip 1em plus 0.5em minus 0.4em\relax IEEE, 2023, pp. 436--443.

\bibitem{yu2023nezha}
G.~Yu, P.~Chen, Y.~Li, H.~Chen, X.~Li, and Z.~Zheng, ``Nezha: Interpretable fine-grained root causes analysis for microservices on multi-modal observability data,'' in \emph{Proceedings of the 31st ACM Joint European Software Engineering Conference and Symposium on the Foundations of Software Engineering}, 2023, pp. 553--565.

\bibitem{rosenberg2020spectrum}
C.~M. Rosenberg and L.~Moonen, ``Spectrum-based log diagnosis,'' in \emph{Proceedings of the 14th ACM/IEEE International Symposium on Empirical Software Engineering and Measurement (ESEM)}, 2020, pp. 1--12.

\bibitem{jia2021logflash}
T.~Jia, Y.~Wu, C.~Hou, and Y.~Li, ``Logflash: Real-time streaming anomaly detection and diagnosis from system logs for large-scale software systems,'' in \emph{2021 IEEE 32nd International Symposium on Software Reliability Engineering (ISSRE)}.\hskip 1em plus 0.5em minus 0.4em\relax IEEE, 2021, pp. 80--90.

\bibitem{shi2023serverrca}
J.~Shi, S.~Jiang, B.~Xu, and Y.~Xiao, ``Serverrca: Root cause analysis for server failure using operating system logs,'' in \emph{2023 IEEE 34th International Symposium on Software Reliability Engineering (ISSRE)}.\hskip 1em plus 0.5em minus 0.4em\relax IEEE, 2023, pp. 486--496.

\bibitem{sultani2018real}
W.~Sultani, C.~Chen, and M.~Shah, ``Real-world anomaly detection in surveillance videos,'' in \emph{Proceedings of the IEEE conference on computer vision and pattern recognition}, 2018, pp. 6479--6488.

\bibitem{midlog}
M.~He, T.~Jia, C.~Duan, H.~Cai, Y.~Li, and G.~Huang, ``Weakly-supervised log-based anomaly detection with inexact labels via multi-instance learning,'' in \emph{2025 IEEE/ACM 47th International Conference on Software Engineering (ICSE)}.\hskip 1em plus 0.5em minus 0.4em\relax IEEE Computer Society, 2025, pp. 726--726.

\bibitem{oliner2007supercomputers}
A.~Oliner and J.~Stearley, ``What supercomputers say: A study of five system logs,'' in \emph{37th annual IEEE/IFIP international conference on dependable systems and networks (DSN'07)}.\hskip 1em plus 0.5em minus 0.4em\relax IEEE, 2007, pp. 575--584.

\bibitem{he2017drain}
P.~He, J.~Zhu, Z.~Zheng, and M.~R. Lyu, ``Drain: An online log parsing approach with fixed depth tree,'' in \emph{2017 IEEE international conference on web services (ICWS)}.\hskip 1em plus 0.5em minus 0.4em\relax IEEE, 2017, pp. 33--40.

\bibitem{misra2016cross}
I.~Misra, A.~Shrivastava, A.~Gupta, and M.~Hebert, ``Cross-stitch networks for multi-task learning,'' in \emph{Proceedings of the IEEE conference on computer vision and pattern recognition}, 2016, pp. 3994--4003.

\bibitem{ruder2019latent}
S.~Ruder, J.~Bingel, I.~Augenstein, and A.~S{\o}gaard, ``Latent multi-task architecture learning,'' in \emph{Proceedings of the AAAI conference on artificial intelligence}, vol.~33, no.~01, 2019, pp. 4822--4829.

\bibitem{lin2020shared}
P.~Lin and M.~Yang, ``A shared-private representation model with coarse-to-fine extraction for target sentiment analysis,'' in \emph{Findings of the Association for Computational Linguistics: EMNLP 2020}, 2020, pp. 4280--4289.

\bibitem{yue2022dare}
L.~Yue, Q.~Liu, Y.~Du, Y.~An, L.~Wang, and E.~Chen, ``Dare: disentanglement-augmented rationale extraction,'' \emph{Advances in Neural Information Processing Systems}, vol.~35, pp. 26\,603--26\,617, 2022.

\bibitem{nakano2023interaction}
A.~Nakano, M.~Suzuki, and Y.~Matsuo, ``Interaction-based disentanglement of entities for object-centric world models,'' in \emph{The Eleventh International Conference on Learning Representations}, 2023.

\bibitem{bousmalis2016domain}
K.~Bousmalis, G.~Trigeorgis, N.~Silberman, D.~Krishnan, and D.~Erhan, ``Domain separation networks,'' \emph{Advances in neural information processing systems}, vol.~29, 2016.

\bibitem{vaswani2017attention}
A.~Vaswani, ``Attention is all you need,'' \emph{Advances in Neural Information Processing Systems}, 2017.

\bibitem{jiang2023weakly}
M.~Jiang, C.~Hou, A.~Zheng, X.~Hu, S.~Han, H.~Huang, X.~He, P.~S. Yu, and Y.~Zhao, ``Weakly supervised anomaly detection: A survey,'' \emph{arXiv preprint arXiv:2302.04549}, 2023.

\bibitem{carbonneau2018multiple}
M.-A. Carbonneau, V.~Cheplygina, E.~Granger, and G.~Gagnon, ``Multiple instance learning: A survey of problem characteristics and applications,'' \emph{Pattern Recognition}, vol.~77, pp. 329--353, 2018.

\bibitem{perini2023learning}
L.~Perini, V.~Vercruyssen, and J.~Davis, ``Learning from positive and unlabeled multi-instance bags in anomaly detection,'' in \emph{Proceedings of the 29th ACM SIGKDD Conference on Knowledge Discovery and Data Mining}, 2023, pp. 1897--1906.

\bibitem{chen2022hierarchical}
W.~Chen, J.~Du, Z.~Zhang, F.~Zhuang, and Z.~He, ``A hierarchical interactive network for joint span-based aspect-sentiment analysis,'' \emph{arXiv preprint arXiv:2208.11283}, 2022.

\bibitem{su2023towards}
W.~Su, X.~Zhu, C.~Tao, L.~Lu, B.~Li, G.~Huang, Y.~Qiao, X.~Wang, J.~Zhou, and J.~Dai, ``Towards all-in-one pre-training via maximizing multi-modal mutual information,'' in \emph{Proceedings of the IEEE/CVF Conference on Computer Vision and Pattern Recognition}, 2023, pp. 15\,888--15\,899.

\bibitem{bachman2019learning}
P.~Bachman, R.~D. Hjelm, and W.~Buchwalter, ``Learning representations by maximizing mutual information across views,'' in \emph{Proceedings of the 33rd International Conference on Neural Information Processing Systems}, 2019, pp. 15\,535--15\,545.

\bibitem{le2021log}
V.-H. Le and H.~Zhang, ``Log-based anomaly detection without log parsing,'' in \emph{2021 36th IEEE/ACM International Conference on Automated Software Engineering (ASE)}.\hskip 1em plus 0.5em minus 0.4em\relax IEEE, 2021, pp. 492--504.

\bibitem{le2022log}
------, ``Log-based anomaly detection with deep learning: How far are we?'' in \emph{Proceedings of the 44th international conference on software engineering}, 2022, pp. 1356--1367.

\bibitem{li2022swisslog}
X.~Li, P.~Chen, L.~Jing, Z.~He, and G.~Yu, ``Swisslog: Robust anomaly detection and localization for interleaved unstructured logs,'' \emph{IEEE Transactions on Dependable and Secure Computing}, 2022.

\bibitem{lu2018detecting}
S.~Lu, X.~Wei, Y.~Li, and L.~Wang, ``Detecting anomaly in big data system logs using convolutional neural network,'' in \emph{2018 IEEE 16th Intl Conf on Dependable, Autonomic and Secure Computing, 16th Intl Conf on Pervasive Intelligence and Computing, 4th Intl Conf on Big Data Intelligence and Computing and Cyber Science and Technology Congress (DASC/PiCom/DataCom/CyberSciTech)}.\hskip 1em plus 0.5em minus 0.4em\relax IEEE, 2018, pp. 151--158.

\bibitem{duan2023afalog}
C.~Duan, T.~Jia, H.~Cai, Y.~Li, and G.~Huang, ``Afalog: A general augmentation framework for log-based anomaly detection with active learning,'' in \emph{2023 IEEE 34th International Symposium on Software Reliability Engineering (ISSRE)}.\hskip 1em plus 0.5em minus 0.4em\relax IEEE, 2023, pp. 46--56.

\bibitem{icsme2020}
K.~Yin, M.~Yan, L.~Xu, Z.~Xu, Z.~Li, D.~Yang, and X.~Zhang, ``Improving log-based anomaly detection with component-aware analysis,'' in \emph{2020 IEEE International Conference on Software Maintenance and Evolution (ICSME)}, 2020, pp. 667--671.

\bibitem{icse2020}
J.~Kim, V.~Savchenko, K.~Shin, K.~Sorokin, H.~Jeon, G.~Pankratenko, S.~Markov, and C.-J. Kim, ``Automatic abnormal log detection by analyzing log history for providing debugging insight,'' in \emph{Proceedings of the ACM/IEEE 42nd International Conference on Software Engineering: Software Engineering in Practice}, ser. ICSE-SEIP '20.\hskip 1em plus 0.5em minus 0.4em\relax New York, NY, USA: Association for Computing Machinery, 2020, p. 71–80.

\bibitem{logflash}
T.~Jia, Y.~Wu, C.~Hou, and Y.~Li, ``Logflash: Real-time streaming anomaly detection and diagnosis from system logs for large-scale software systems,'' in \emph{2021 IEEE 32nd International Symposium on Software Reliability Engineering (ISSRE)}, 2021, pp. 80--90.

\bibitem{kddcfg}
A.~Nandi, A.~Mandal, S.~Atreja, G.~B. Dasgupta, and S.~Bhattacharya, ``Anomaly detection using program control flow graph mining from execution logs,'' in \emph{Proceedings of the 22nd ACM SIGKDD International Conference on Knowledge Discovery and Data Mining}, ser. KDD '16.\hskip 1em plus 0.5em minus 0.4em\relax New York, NY, USA: Association for Computing Machinery, 2016, p. 215–224.

\bibitem{logsed}
T.~Jia, L.~Yang, P.~Chen, Y.~Li, F.~Meng, and J.~Xu, ``Logsed: Anomaly diagnosis through mining time-weighted control flow graph in logs,'' in \emph{2017 IEEE 10th International Conference on Cloud Computing (CLOUD)}, 2017, pp. 447--455.

\bibitem{icwslogsed}
T.~Jia, P.~Chen, L.~Yang, Y.~Li, F.~Meng, and J.~Xu, ``An approach for anomaly diagnosis based on hybrid graph model with logs for distributed services,'' in \emph{2017 IEEE International Conference on Web Services (ICWS)}, 2017, pp. 25--32.

\bibitem{zhang2021survey}
Y.~Zhang and Q.~Yang, ``A survey on multi-task learning,'' \emph{IEEE transactions on knowledge and data engineering}, vol.~34, no.~12, pp. 5586--5609, 2021.

\bibitem{standley2020tasks}
T.~Standley, A.~Zamir, D.~Chen, L.~Guibas, J.~Malik, and S.~Savarese, ``Which tasks should be learned together in multi-task learning?'' in \emph{International conference on machine learning}.\hskip 1em plus 0.5em minus 0.4em\relax PMLR, 2020, pp. 9120--9132.

\bibitem{vandenhende2021multi}
S.~Vandenhende, S.~Georgoulis, W.~Van~Gansbeke, M.~Proesmans, D.~Dai, and L.~Van~Gool, ``Multi-task learning for dense prediction tasks: A survey,'' \emph{IEEE transactions on pattern analysis and machine intelligence}, vol.~44, no.~7, pp. 3614--3633, 2021.

\bibitem{long2017learning}
M.~Long, Z.~Cao, J.~Wang, and P.~S. Yu, ``Learning multiple tasks with multilinear relationship networks,'' \emph{Advances in neural information processing systems}, vol.~30, 2017.

\bibitem{bu2021asap}
J.~Bu, L.~Ren, S.~Zheng, Y.~Yang, J.~Wang, F.~Zhang, and W.~Wu, ``Asap: A chinese review dataset towards aspect category sentiment analysis and rating prediction,'' \emph{arXiv preprint arXiv:2103.06605}, 2021.

\bibitem{lee2021private}
M.~Lee and V.~Pavlovic, ``Private-shared disentangled multimodal vae for learning of latent representations,'' in \emph{Proceedings of the ieee/cvf conference on computer vision and pattern recognition}, 2021, pp. 1692--1700.

\bibitem{he2019interactive}
R.~He, W.~S. Lee, H.~T. Ng, and D.~Dahlmeier, ``An interactive multi-task learning network for end-to-end aspect-based sentiment analysis,'' in \emph{Proceedings of the 57th Annual Meeting of the Association for Computational Linguistics}, 2019, pp. 504--515.

\bibitem{liang2023stage}
S.~Liang, W.~Wei, X.-L. Mao, Y.~Fu, R.~Fang, and D.~Chen, ``Stage: span tagging and greedy inference scheme for aspect sentiment triplet extraction,'' in \emph{Proceedings of the AAAI Conference on Artificial Intelligence}, vol.~37, no.~11, 2023, pp. 13\,174--13\,182.

\end{thebibliography}

\end{document}